\documentclass[bibyear]{aa}
\usepackage{graphicx}
\usepackage{amssymb}
\usepackage{tabulary}
\usepackage{natbib,twoopt}
\usepackage{placeins}

\begin{document}
   \title{Studying the gaseous outskirts of galaxy groups \\
   with coherent Ly\,$\alpha$ absorption patterns
%  \subtitle{}
   \thanks{Based on observations obtained with the NASA/ESA
   Hubble Space Telescope, which is operated by the Space
   Telescope Science Institute (STScI) for the Association of
   Universities for Research in Astronomy, Inc., under NASA
   contract NAS5D26555.}}

   \author{
          P. Richter \inst{1},
          J.C. Charlton  \inst{2},
          A.J. Fox \inst{3},
          Sameer \inst{4,2}
          \and
          B.P. Wakker \inst{5}
          }
   \offprints{P. Richter\\
   \email{prichter@astro.physik.uni-potsdam.de}}

   \institute{Institut f\"ur Physik und Astronomie, Universit\"at Potsdam,
             Karl-Liebknecht-Str.\,24/25, 14476 Golm, Germany
   \and
   Department of Astronomy \& Astrophysics, 525 Davey Lab, The Pennsylvania State University, \\
   University Park, PA 16802, USA
   \and
   AURA for ESA, Space Telescope Science Institute, 3700 San Martin Drive,
   Baltimore, MD 21218, USA
   \and
   Department of Physics \& Astronomy, Nieuwland Science Hall, The University of Notre Dame, \\
   Notre Dame, IN 46556, USA
   \and
   Eureka Scientific, Inc., 2452 Delmer Street, Oakland, CA 94602, USA
   }

   \date{Received 23 November 2024; accepted 15 July 2025}

%%%%%%%%%%%%%%%%%%%%%% ABSTRACT %%%%%%%%%%%%%%%%%%%%%%

\abstract
{}
{
In this study, we explore the properties of diffuse intergalactic gas residing in
the outskirts of the four nearby, low-mass galaxy groups NGC\,1052, NGC\,5866, NGC\,4631,
and NGC\,3992 (all at $cz\leq 2000$ km\,s$^{-1}$) beyond their group virial radii.
}
{
Using archival ultraviolet absorption spectra of bright active galactic nuclei (AGN)
observed with the {\it Hubble Space Telescope} (HST) and its Cosmic Origins Spectrograph (COS),
we search for H\,{\sc i} Ly\,$\alpha$ absorption near the group's recession
velocities along 35 sightlines that pass the outer group medium (OGrM) at
normalized impact parameters $\rho /R_{\rm vir}=1-3$.
We derive H\,{\sc i} column densities of the absorbers and constrain the physical
conditions in the gas (thermal pressure, density, neutral gas fraction, absorption path-length)
by using a hydrostatic toy model of the group's gas environment and assuming photoionization.
}
{
H\,{\sc i} Ly\,$\alpha$ absorption near the group's recession velocities is
detected along 19 sightlines with H\,{\sc i} column densities in the range
log $(N$(H\,{\sc i})/cm$^{-2})=12.50-14.34$, implying a high OGrM detection rate
of more than $50$ percent.
We transform this value into an incidence rate of OGrM absorbers per unit redshift of
$d{\cal N}/dz=232 \pm 58$ for absorbers with log $(N$(H\,{\sc i})/cm$^{-2})\geq 13.2$ and
$\rho /R_{\rm vir}=1-3$.
This is $25$ percent above the value derived for the general population of Ly\,$\alpha$ absorbers
within $z=0$ filaments and more than twice the value for the $z=0$ Ly\,$\alpha$ forest
(considering the same column-density range). From the modeling, we obtain
lower limits for the gas densities from log ($n_{\rm H}/$cm$^{-3})=-5.00$ to $-3.72$, comparable
to densities found in the overall Ly\,$\alpha$ forest.
}
{
Our study unveils a large cross section and overdensity of Ly\,$\alpha$ absorbers in the
outskirts of these four nearby groups. Such an overdensity is in line with a previously
proposed scenario, in which AGN feedback lifts gaseous material to large distances beyond
the virial radius of groups into the OGrM.
However, a larger survey of OGrM absorbers and a comparison with hydrodynamical simulations
will be necessary to constrain the cosmological mass density of OGrM absorbers and pinpoint their
role in cosmological structure formation and galaxy/group evolution.
}

\titlerunning{Properties of the OGrM in four nearby groups}

\maketitle

%
%________________________________________________________________

%%%%%%%%%%%%%%%%%%%%%% TABLE 01 %%%%%%%%%%%%%%%%%%%%%%

\begin{table*}[t!]
\caption[]{Group parameters}
\begin{scriptsize}
\begin{tabular}{llrrrrrrrrr}
\hline
Group name & Alt.\,group names$^{\rm a}$ &
$\alpha_{2000}$$^{\rm b}$ & $\delta_{2000}$$^{\rm b}$ &  $D$$^{\rm c}$ & $v_{\rm gr}$$^{\rm d}$ & 
$N_{\rm gal}$$^{\rm e}$ & log & log & $R_{\rm vir}$ & $\sigma_v$$^{\rm h}$ \\
 & & [deg] & [deg] & [Mpc] & [km\,s$^{-1}$] & & $(M_{\rm dyn}/M_{\sun})$$^{\rm f}$ & 
$(M_{L}/M_{\sun})$$^{\rm g}$ & [Mpc] & [km\,s$^{-1}$] \\
\hline
\hline
NGC\,1052 group & LGG\,71            &  40.107 & $-$7.883 & 19 & $1240-1540$ &  9 & 12.42 & 12.66 & 0.45 & 105.7 \\
NGC\,5866 group & LGG\,396           & 228.368 & 56.250   & 16 &  $650-850$  &  5 & 12.09 & 12.34 & 0.35 &  82.0 \\
NGC\,4631 group & LGG\,291, Coma\,I  & 187.645 & 30.372   &  9 &  $520-820$  & 12 & 12.36 & 12.62 & 0.43 & 100.5 \\
NGC\,3992 group & UMaI\,N, LGG\,258  & 177.431 & 53.648   & 20 &  $640-1280$ & 32 & 13.03 & 13.27 & 0.72 & 168.3 \\
\hline
\end{tabular}
\noindent
\\
$^{\rm a}$\,see, e.g., \citet{fouque1992} and \citet{garcia1993};
$^{\rm b}$\,luminosity-weighted group center coordinate;
$^{\rm c}$\,distance adopted from NED;
$^{\rm d}$\,group's radial velocity range (see Sect.\,2.2);
$^{\rm e}$\,number of bright member galaxies \citep[from][]{fouque1992}.
$^{\rm f}$\,dynamical mass;
$^{\rm g}$\,luminosity-based mass;
$^{\rm h}$\,velocity dispersion from group member galaxies.\\
\end{scriptsize}
\end{table*}

%%%%%%%%%%%%%%%%%%%%%% SECTION 01 %%%%%%%%%%%%%%%%%%%%%%

\section{Introduction}

Galaxy groups in the mass range $M=10^{12}-10^{14} M_{\sun}$ represent particularly representative
cosmological structures in the Universe, as they harbor the majority of the star-forming galaxies 
in the Universe \citep{mulchaey2000,eke2006,lovisari2021}. In addition, the Dark Matter (DM) halo mass
function at low redshift peaks in this mass range \citep[e.g.,][]{despali2014}. In the canonical picture
of the Universe's large-scale structure, such low-mass galaxy group environments represent
mildly overdense matter knots in the filamentary Cosmic Web \citep{tempel2014}.
In contrast to more massive groups and clusters, the intra-group medium (IGrM) in low-mass groups 
is too cold (and too diffuse) to be detectable in X-ray emission throughout the overall group 
environment \citep{mulchaey1996}. This leaves absorption-line spectroscopy against bright AGN as the only 
efficient method to systematically explore the diffuse, multi-phase IGrM with a particular focus on 
H\,{\sc i} Ly\,$\alpha$/Ly\,$\beta$ and, potentially, low, intermediate and high metal ions such 
as C\,{\sc ii}, Mg\,{\sc ii}, Si\,{\sc iii}, Si\,{\sc iv}, C\,{\sc iv}, and O\,{\sc vi} 
\citep{savage2014,stocke2014,richter2016,bielby2017,sameer2022}. Recent absorption-line studies of individual group 
environments unveil absorber properties and absorber/galaxy relations that differ from those 
in isolated galaxy environments \citep{stocke2019,pointon2017,pointon2020}, reflecting the on-going 
cosmological structure formation in these overdense regions. Star-formation activity, triggered by 
gas-accretion processes in groups, will lead to outflows and winds, depositing energy 
and heavy elements in the IGrM that can be detected by measuring high-ion absorption (e.g., from
C\,{\sc iv} and O\,{\sc vi}) along sightlines that pass group environments 
\citep[e.g.,][]{burchett2018,stocke2019,sameer2022}.
Other case studies of pre-selected group environments with multi-wavelength observations 
demonstrate that galaxy mergers have a crucial influence on the 
large-scale properties of the IGrM \citep{yun1994,borthakur2010,richter2018}. This is supported 
by zoom-in cosmological simulations of groups, which demonstrate that the covering 
fraction of high-column density gas in the IGrM rises substantially in the time-frame of merger 
events, leaving their imprints in the cross-section of intermediate and high ions (see 
Oppenheimer et al.\,2021 and references therein). Characterizing the gas-circulation processes 
in the IGrM around galaxies as a function of their immediate cosmological environment therefore
is of fundamental importance to understand cosmological structure-formation. 

%%%%%%%%%%%%%%%%%%%%%%% FIGURE 01 %%%%%%%%%%%%%%%%%%%%%%%

\begin{figure*}[ht!]
\begin{center}
\resizebox{0.86\hsize}{!}{\includegraphics{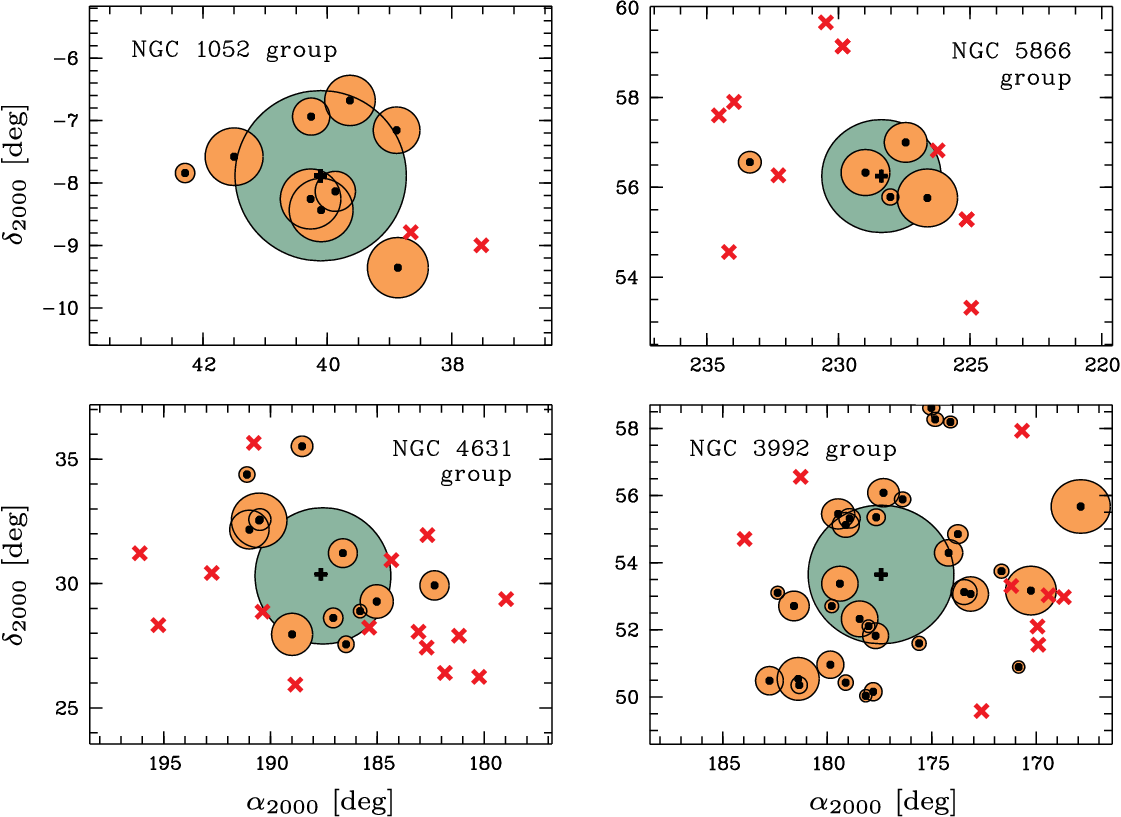}}
\caption[]{
Spatial distribution of the galaxies in the NGC\,1052, NGC\,5866, NGC\,4631, NGC\,3992
galaxy groups together with the galaxies' (orange-shaded circles) and groups' (green-shaded
circles) virial radii and the position of the selected background AGN sampled by
HST/COS data (red crosses).
}
\end{center}
\end{figure*}

%%%%%%%%%%%%%%%%%%%%%%%%%%%%%%%%%%%%%%%%%%%%%%%%%%%%%%%%%

So far, very little attention has been paid in the literature to absorption signatures of gas 
in the outskirts of low-mass galaxy groups beyond the group's virial radius
(and beyond the virial radius of group member galaxies), where the groups connect to the Cosmic Web.
As recent studies imply, the gaseous outskirts of groups are believed to host a substantial 
amount of baryons in the form of warm/hot gas, a direct result of AGN feedback that potentially
lifts gaseous material to large distances even beyond the group's virial radii 
\citep[see][and references therein]{eckert2021}. 
From this reservoir, group member galaxies can accrete metal-poor gas to 
fuel star formation, making this gas component particularly interesting for our understanding
of galaxy evolution.
In this interface region between the IGrM and the ambient filamentary IGM, the mean gas density 
is expected to be very low (close to that of the IGM itself) where $n_{\rm H}<10^{-4}$ cm$^{-3}$, 
typically \citep{martizzi2019}. As a result, metal absorption is expected to be 
very rare, leaving H\,{\sc i} Ly\,$\alpha$ absorption as the only sensitive tracer of this 
gas, which we hereafter refer to as outer-group medium (OGrM). 
Note that, in this definition, OGrM Ly\,$\alpha$ absorbers represent a sub-class of 
Ly\,$\alpha$ filament absorbers \citep[e.g.,][]{wakker2015,bouma2022} and the overall
Ly\,$\alpha$ forest \citep[e.g.,][]{lehner2007,danforth2016}, as groups and their gaseous 
environment represent galaxy overdensities within the hierarchically-structured Cosmic Web.
As the OGrM is expected to be gravitationally coupled to the group's DM potential, 
OGrM absorbers are expected to cluster around the group's recession velocity/redshift. This
is possibly separating them from ``free-floating'' absorption systems within the Cosmic Web
or in voids that do not have such a clear kinematic association with galaxy concentrations.

The above considerations motivated us to start a systematic search for OGrM H\,{\sc i}
Ly\,$\alpha$ absorption features in archival HST/COS spectra along sightlines that pass
nearby galaxy groups and analyze their properties. In this pilot study, we present our 
results for four nearby galaxy groups at recession velocities $cz\leq 2000$ km\,s$^{-1}$:
the NGC\,1052 group, the NGC\,5866 group, the NGC\,4631 group, and the NGC\,3992 group.

This paper is organized as follows: 
in Sect.\,2, we present the search strategy, the data basis and handling and the 
analysis method. In Sect.\,3, we discuss the individual OGrM absorbers and
their relation to the groups. The properties of the absorbers and the 
physical conditions in the gas are discussed in the Sect.\,4. 
A summary of our study as well as final conclusions are presented in Sect.\,5.
The appendix provides additional information on the continuum-fitting process
and the hydrostatic toy model.

%%%%%%%%%%%%%%%%%%%%%% SECTION 02 %%%%%%%%%%%%%%%%%%%%%%

\section{Data handling and analysis method}

\subsection{Search strategy}

To study diffuse gas in the large-scale environment of nearby galaxy groups, we 
have combined archival ultraviolet (UV) spectral data from HST/COS with publicly available galaxy 
data and information from various galaxy-group surveys.

As spectral data basis, we used 1040 COS G130M+G160M data sets of distant ($z>0.01$) 
AGN and galaxies downloaded from the MAST archive, as available to us by
February 2025. For this, we used the data selection 
strategy outlined in our previous HST/COS archival surveys \citep{richter2016,richter2017}.
Since we are aiming at studying the H\,{\sc i} Ly\,$\alpha$ absorption signatures of 
very nearby (in a cosmological sense) galaxy groups at recession velocities 
$cz\leq 5000$ km\,s$^{-1}$, we have pre-selected those 311 AGN spectra that have a  
signal-to-noise ratio of S/N$\geq 6$ per 17 km\,s$^{-1}$ wide spectral resolution 
element of COS at $1230$ \AA. This S/N cutoff was chosen because of the fact
that it becomes very difficult to unambiguously identify weak absorption features
in the extended wing of the Galactic interstellar Ly\,$\alpha$ absorption at S/N
ratios $<6$. We also do not consider those spectra, for which the flux distribution
in the red wing of the interstellar Ly\,$\alpha$ absorption is highly irregular 
(so that a reliable reconstruction of the continuum is not possible) or that
exhibit a critical number of blending lines at $\lambda=1220-1230$ \AA\,
from other systems at higher redshift.

To identify $z\approx 0$ galaxy groups that lie in the vicinity of the pre-selected archival 
HST/COS AGN sightlines, we used the classical group catalog of very nearby 
($cz\leq 6000$ km\,s$^{-1}$) galaxy groups and their brightest member galaxies from
\citet{fouque1992} to calculate the group's mass and radial extent (see Sect.\,2.2).
For each group listed in this catalog, we created finding charts by cross-correlating the 
positions of the COS AGN with the group's positions in an impact-parameter range 
between $1-3\,R_{\rm vir}$. 
In this way, we identified outer-group environments at $cz\leq 5000$ km\,s$^{-1}$ 
that are sampled by at least two AGN sightlines. All in all, we have identified 
14 galaxy groups in this manner whose OGrM can be studied in H\,{\sc i} Ly\,$\alpha$ absorption
with archival HST/COS data.

In this study, we focus on four particularly interesting nearby group environments 
at $cz\leq 2000$ km\,s$^{-1}$: the NGC\,1052 group, the NGC\,5866 group, the NGC\,4631 group, 
and the NGC\,3992 group. Throughout the following, we use the group names from the original work by 
\citet{fouque1992} except for group labeled as "Ursa Major\,I North group" in 
\citet{fouque1992}, for which we use the alternative name
``NGC\,3992 group'' to avoid confusion with the labeling of Ursa Major group environments used
in other group catalogues \citep[see][]{wolfinger2016}. 
The ``NGC\,4631 group'' is also known
under the name ``Coma I group'' and the circumgalactic environment of NGC\,4631 itself
previously has been studied by us using UV absorption-line data from HST/COS in combination
with H\,{\sc i} 21cm from the HALOGAS survey \citep{richter2018}.
Alternative group names are listed in Table 1, second column.

Based on our HST/COS archival data set and the above-mentioned selection criteria,
we identify 56 HST/COS spectra that sample the outer environment of these four
groups in the range $\rho /R_{\rm vir}=1-3$. From these, however, 21 sightlines exhibit Ly\,$\alpha$
absorption that we associate with the circumgalactic medium (CGM) of group-member galaxies or 
foreground galaxies, as discussed in Sect.\,2.3. Therefore, our final, CGM-corrected spectral sample consists 
of 35 AGN sightlines that can be used to study the OGrM of these four groups.

Some of these AGN have been observed by COS at a relatively high S/N of $>20$ per resolution 
element, so that very weak H\,{\sc i} Ly\,$\alpha$ absorbers at column densities
$N$(H\,{\sc i}$)\leq 10^{13}$ cm$^{-2}$ can also be detected. These four groups, together with 
the rich HST/COS data set, therefore represent an ideal starting 
point for a sensitive first systematic study of the OGrM in Ly\,$\alpha$ absorption. 
In Table 1, we summarize the main properties of the NGC\,1052, NGC\,5866, NGC\,4631, 
and NGC\,3992 groups.

\subsection{Group parameters and galaxy data}

For our study, we use the group catalog from \citet{fouque1992}, which is 
particularly strict in defining group members. In this catalog, only the 
largest group member galaxies with isophotal diameters $D_{25}\geq 100$ 
\footnote{$D_{25}$ is the isophotal diameter measured at a surface brightness
of 25 magnitudes per square arcsecond.} arcsec are considered, 
which contain most of the stellar mass and luminosity of these groups. As described in 
\citet{gourgoulhon1992}, the group members were identified using a hierarchical 
finding algorithm, where the hierarchy is built on the mass density of the 
aggregates progressively formed by the algorithm. Only those galaxy aggregates 
are considered that have a B-band luminosity density higher than 
$8\times 10^9\,L_{\sun}$\,Mpc$^{-3}$, ensuring that the group is gravitationally bound.
In addition, the galaxy-selection criterion, together with the hierarchical method, warrants 
that contamination with field galaxies and due to projection effects with filaments 
is minimized \citep[see discussion in][]{gourgoulhon1992}.
The latter two aspects are particularly important for our study, as we use the
kinematic information on the confirmed members galaxies to define the search window for
Ly\,$\alpha$ absorption associated with the groups and separate the OGrM absorber
statistics from that of ``field''/filament Ly\,$\alpha$ absorbers \citep{danforth2016,bouma2022,wakker2015}.

Galaxy data for the 58 confirmed group members \citep[as listed in][see Table 1, 7th column]{fouque1992}
and for other, potential group member galaxies that are spatially and kinematically aligned with the four 
groups have been extracted from the NED and SIMBAD online data bases. These data include J$2000$ 
coordinates, B-band magnitudes, distances, and radial velocities for a total of 430 galaxies in the 
four group fields displayed in Fig.\,1. Since all these galaxies are very nearby ($D\leq 20$ Mpc),
we estimate that our galaxy sample is complete down to luminosities of $\sim 0.08\,L^{\star}$
\citep[see][their Fig.\,1]{french2017}.

For the group member galaxies, we have derived B-band luminosities and virial radii ($R_{\rm vir}$) 
of the galaxies using the scaling relation from \citet[][their equation (3)]{richter2016}, which 
is based on the approach described in \citet{stocke2013}.
From the positions of the galaxies, we have then calculated the luminosity-weighted group center 
coordinates, as given in Table 1 (columns 3 \& 4). For each of the four groups, we then derived the 
velocity dispersion of the group member galaxies using the relation from \citet{osmond2004},

\begin{equation}
\sigma_v = \sqrt{\frac{\Sigma\,(v-\bar{v})^2}{N_{\rm gal}-3/2}}
\pm \frac{\sigma_v}{\sqrt{2(N_{\rm gal}-3/2)}},
\end{equation}

where $N_{\rm gal}$ denotes the number of group members, $v$ is the velocity of each individual group member, 
and $\bar{v}$ is the mean galaxy velocity in the group. For the determination of the {\it dynamical}
mass of each group (Table 1, 8th column), we have used the relation of \citet{tully2015}, 

\begin{equation}
\frac{M_{\rm dyn}}{M_{\sun}} = 
1.5\times 10^6\,h^{-1}\,\left(\frac{\sigma_v}{{\rm km\,s}^{-1}}\right)^3,
\end{equation}

where we have to assume the groups to be virialized. For low-mass groups,
however, the latter assumption may not hold \citep[e.g.][]{marini2024},
introducing a substantial systematic uncertainty in the determination of 
the dynamical group mass.

An alternative way to determine the mass of low-mass groups
is to consider the group's total luminosity \citep[see discussion in][]{marini2024}.
To do this, we use the luminosity-to-mass conversion scheme presented in 
\citet[][their equation (4)]{stocke2019} together with a group mass-to-light
ratio of $\curlyvee_{\rm grp}=M_{\rm grp}/L_{\rm grp}=150$, as appropriate for 
groups with total luminosities log $(L_{\rm grp}/L_{\sun})\leq 11.2$
\citep{proctor2015}, to define

\begin{equation}
\frac{M_L}{M_{\sun}} =
\curlyvee_{\rm grp}\,10^{10}\,\left(\frac{L_{\rm grp}}{L^{\star}}\right).
\end{equation}

The luminosity-based group masses, $M_L$, derived in this way are listed in Table 1, 
9th column. As can be seen, the values for $M_L$ are slightly higher than for
$M_{\rm dyn}$ ($\sim 0.25$ dex), possibly related to the fact that these groups
are not (yet) fully virialized. For the following, we therefore use $M_L$ as a measure
for the groups' total masses.

To characterize the groups' radial extent, we have calculated $R_{\rm vir}$ for each group 
under the assumption that that their potentials follow
an NFW profile \citep{navarro1995}. For this, we have used the relation 
(A1) given in \citet{richter2020}, which is based on the work by \citet{maller2004},
where we assume a virial overdensity of $\Delta_{\rm vir}=360$ at the virialization 
redshift of $z=0$ \citep[see also][]{shull2014}.
\footnote{Note that in the context of galaxy groups and clusters, their
characteristic radius often is expressed in terms of $r_{200}$,
the radius for which the mass density profile of a halo
equals 200 times the critical density of the Universe; at $z=0$,
$R_{\rm vir} > r_{200}$ and $R_{\rm vir} \approx r_{100}$
\citep{bryan1998}.}

While we do not use the data for the other galaxies in the group fields to determine
group masses and virial radii (as their group-membership status is unclear),
we do consider these galaxies in Sect.\,3 to identify (and sort out) CGM absorbers that
arise in the halos of these ``field'' galaxies.

In Fig.\,1, we display the spatial distribution of the galaxies in the four
groups together with the member galaxies' $R_{\rm vir}$ and the groups' 
virial radii. The positions of the background AGN are indicated with red crosses.

\subsection{COS data and their spectral analysis}

The overall strategy of the data reduction and analysis method of the HST/COS data follows 
the approach that we have used in our previous surveys and studies involving HST/COS
spectral data \citep[e.g.,][]{richter2016,richter2017,richter2020}. The original (raw) data 
of the individual COS science exposures (from the G130M and G160M gratings) of the 
AGN sample available to us were processed by the CALCOS pipeline (v3.4.3) and transformed
into standard x1d fits files. In a second step, the individual exposures then were 
coadded using a custom-written code that aligns individual exposures based on a 
pixel/wavelength calibration \citep[see][for a detailed description]{richter2017}.

For our study of H\,{\sc i} Ly\,$\alpha$ absorption in the OGrM in nearby groups at 
$cz\leq 2000$ km\,s$^{-1}$, the relevant wavelength range is $1215.67-1223.78$ \AA.
This short wavelength interval is covered by the COS G130M grating, which operates 
from $1150-1450$ \AA\,at a spectral resolution of $R\approx 15,000-18,000$ ($\approx 20-17$ 
km\,s$^{-1}$ FWHM), depending on wavelength, central wavelength setting, and the lifetime 
position of the COS detector. The native pixel size in the data is $3$ km\,s$^{-1}$ 
\citep{green2012,debes2016}. With this targeted wavelength range,
the OGrM Ly\,$\alpha$ absorption is expected to arise in the red wing of the damped 
interstellar Ly\,$\alpha$ absorption trough (centered near 1215.67 \AA), 
which typically reaches out to $\lambda =1220-1227$ \AA\, depending on the local 
H\,{\sc i} column density in the direction of the background AGN. Therefore, each 
COS spectrum used in this study was carefully continuum normalized in the $1215-1230$ \AA\, 
range using the custom-written spectral analysis code {\tt span} \citep{richter2013} together 
with a Voigt-profile model of the foreground Milky Way interstellar Ly\,$\alpha$ absorption. 
An example for the continuum reconstruction of the interstellar Ly\,$\alpha$ trough
towards NGC\,985 is presented in the Appendix in Fig.\,A.1. 

As mentioned above, a significant fraction of the sightlines in our original sample of 56 HST/COS 
spectra passes the immediate environment of group member galaxies or foreground galaxies,
possibly tracing their CGM rather than the OGrM of the group. As criterion for a positive 
CGM Ly\,$\alpha$ detection, we require that the sightline passes a galaxy at an impact 
parameter $\rho\leq R_{\rm vir}$ with the absorption arising within $\pm 500$ km\,s$^{-1}$ of 
the galaxy's systemic velocity. We find 21 such cases in our sample and remove these
sightlines from our analysis, which leaves 35 sightlines for the OGrM analysis (Table 2).
Note that sightlines that do pass a galaxy at  $\rho\leq R_{\rm vir}$ and that do not
exhibit Ly\,$\alpha$ absorption in the galaxy's or group's velocity range still 
indicate a significant non-detection for OGrM absorption in our sample.

In the next step, we scanned the remaining 35 COS spectra for intervening absorption features 
that potentially could represent H\,{\sc i} Ly\,$\alpha$ absorption arising in the extended OGrM 
of these groups. For this, we consider a velocity range $\Delta v_{\rm abs} = \Delta v_{\rm gal}\pm 150$
km\,s$^{-1}$, where $\Delta v_{\rm gal}$ is the observed velocity range of the group member
galaxies listed in Table 1, sixth column. Ly\,$\alpha$-forest absorbers that are unrelated 
to the group and its environment could also coincidentally fall in this velocity range, however.
This aspect needs to be taken into account because the line density of the Ly\,$\alpha$ forest at
$z=0$ is substantial, the number of Ly\,$\alpha$ absorbers per unit redshift 
$d{\cal N}/dz$ being $\approx 100$ for log $N$(H\,{\sc i}$)\geq 13.2$ \citep{lehner2007,danforth2016}. 
Since $v=cz$ at $z=0$, this line density implies that the likelihood of a randomly placed Ly\,$\alpha$ 
absorber in a low-$z$ AGN spectrum to coincidentally fall into a pre-defined velocity interval of width
$\Delta v=500$ km\,s$^{-1}$ is non-negligible, namely $\sim 0.17$. The likelihood for a 
double-coincidence absorption in the same velocity interval along two adjacent lines of sight (LOS) 
is much smaller, however, $\sim 0.03$. To minimize the contamination by coincidental Ly\,$\alpha$ absorbers 
in our OGrM sample, we therefore consider Ly\,$\alpha$ absorbers within $\Delta v_{\rm abs}$
as OGrM candidates only if i) they have at least one neighboring sightline with Ly\,$\alpha$ 
absorption in the same group velocity interval and ii) their absorption-velocity ranges overlap with each other. 
In this way, we expect OGrM absorbers to form a Ly\,$\alpha$ 
``coherence pattern'' along adjacent sightlines (as they align in position and velocity, similar as the 
group member galaxies). Similar considerations on random and non-random coincidences of Ly\,$\alpha$ absorbers 
have been used to study the coherence length of IGM structures against AGN pairs 
\citep[e.g.][]{dinshaw1997}.

For each detected OGrM Ly\,$\alpha$ candidate line, we measured the radial velocity and 
equivalent width (EW) using the {\tt span} code and cross-checked 
whether it could be a metal line or higher H\,{\sc i} Lyman-series lines from other 
intervening absorption systems at higher redshift. We did not find, however, convincing
alternative identifications for any of the detected absorption features, supporting the
conclusion that they indeed represent H\,{\sc i} Ly\,$\alpha$ absorption lines.

For the subsequent analysis, we used the component-modeling method implemented in {\tt span} 
to derive column densities, log $N$(H\,{\sc i}), and Doppler 
parameters ($b$ values) for each resolved absorption component. In Fig.\,A.1, we show 
as an example the modeled Ly\,$\alpha$ OGrM absorber at $1222.6$ \AA\,towards NGC\,985.
For those absorbers, for 
which the absorber shape is uncertain (either due to particularly weak or particularly strong
absorption), we apply the apparent optical depth (AOD) method \citep{savage1991} to 
obtain log $N$(H\,{\sc i}) (or limits thereof) from a direct pixel integration 
(without gaining information on $b$). 
Throughout the paper, we give column densities in units [cm$^{-2}$], particle densities
in units [cm$^{-3}$], and temperatures in units [K]. All results from the absorption-line 
measurements and component modeling are summarized in Table 3.

The main properties of the 35 background AGN and the S/N in the COS data are listed in Table 2.
Instead of using the AGN names used in the COS archive, we here list the most appropriate
AGN names (based on the NED/SIMBAD entries) in the second column of Table 2 with alternative names
in parentheses.

%%%%%%%%%%%%%%%%%%%%%% TABLE 02 %%%%%%%%%%%%%%%%%%%%%%

\begin{table*}[t!]
\caption[]{List of AGN sightlines tracing OGrM absortption in the four groups}
\begin{scriptsize}
\begin{tabular}{lllrrrr}
\hline
Group & AGN name(s)$^{\rm a}$ & Type$^{\rm a}$ & $z_{\rm QSO}$\,$^{\rm a}$ & $\alpha_{2000}$\,$^{\rm a}$ [deg] &
$\delta_{2000}$\,$^{\rm a}$ [deg] & S/N$^{\rm b}$\\
\hline
\hline
NGC\,1052 & NGC\,985                                               & Seyfert & 0.04314 &  38.657842 &  $-$8.788061 & 57 \\
          & Mrk\,1044                                              & Seyfert & 0.01645 &  37.523011 &  $-$8.998113 & 50 \\
\hline
NGC\,5866 & SBS\,1537+577 (VV\,487, MCG+10-22-028)                 & Seyfert & 0.07342 & 234.541864 & $+$57.603645 & 10 \\
          & Mrk\,486 (SBS\,1535+547)                               & Seyfert & 0.03900 & 234.160005 & $+$54.559226 & 10 \\
          & Mrk\,290 (SBS\,1534+580)                               & Seyfert & 0.02958 & 233.968346 & $+$57.902643 & 37 \\
          & SBS\,1527+564 (RBS\,1503,LEDA\,2531253)                & Seyfert & 0.09900 & 232.281067 & $+$56.268536 & 13 \\
          & SBS\,1520+598 (SBS\,1521+598,2MASSI\,J1521537+594019)  & Seyfert & 0.28620 & 230.474209 & $+$59.672238 & 10 \\
          & SBS\,1518+593 (RBS\,1483,LEDA\,2816140)                & Seyfert & 0.07810 & 229.840292 & $+$59.139922 & 21 \\
          & SBS\,1503+570 (8C\,1503+570,RX\,J1504.8+5649)          & Seyfert & 0.35871 & 226.231517 & $+$56.822315 & 14 \\
          & SBS\,1459+554 (RX\_J1500.5+5517)                       & Seyfert & 0.40534 & 225.127784 & $+$55.285886 & 15 \\
          & SBS\,1458+534 (LEDA\,2437690)                          & Seyfert & 0.33800 & 224.956648 & $+$53.319188 & 11 \\
\hline
NGC\,4631 & SDSS\,J130429.03+311308.2 (CBS\,339)                   & Quasar  & 0.80561 & 196.120995 & $+$31.218966 & 16 \\
          & SDSS\,J130100.86+281944.7 (A2\,330,CSO\,786)           & Quasar  & 1.36102 & 195.253615 & $+$28.329095 & 13 \\
          & Ton\,133 (CSO\,174)                                    & Quasar  & 0.65228 & 192.751299 & $+$30.428300 & 15 \\
          & SDSS\,J124307.57+353907.1 (CSO\,915,LEDA\,42770)       & Quasar  & 0.54677 & 190.781547 & $+$35.651971 & 17 \\
          & Ton\,635 (SDSS\,J124129.64+285212.0)                   & Quasar  & 0.58910 & 190.373532 & $+$28.869996 & 12 \\
          & SDSS\,J123521.58+255613.5 (2XMM\,J123521.6+255613)     & Seyfert & 0.23993 & 188.839912 & $+$25.937079 & 10 \\
          & WCom (7C\,1219+2830)                                   & BL\,Lac & 0.10289 & 185.382044 & $+$28.232917 &  7 \\
          & RBS\,1090 (CBS\,56,FBQS\,J1217+3056 )                  & Seyfert & 0.30010 & 184.339236 & $+$30.941888 &  6 \\
          & RX\,J1212.2+2803 (2MASS\,J12121725+2803499)            & Seyfert & 0.16758 & 183.071876 & $+$28.063891 & 12 \\
          & RX\,J1210.7+2725 (SDSS\,J121045.64+272536.4)           & Seyfert & 0.23010 & 182.690154 & $+$27.426816 & 15 \\
          & RX\,J1210.6+3157 (7C\,1208+3213)                       & Seyfert & 0.38891 & 182.656554 & $+$31.951672 &  9 \\
          & SDSS\,J120720.99+262429.1 (QSO\,J1207+2624)            & Quasar  & 0.32300 & 181.837400 & $+$26.408100 & 10 \\
          & PG\,1202+281 (LEDA\,38224)                             & Seyfert & 0.16530 & 181.175457 & $+$27.903297 &  9 \\
          & RX\,J1200.9+2615 (SDSS\,J120056.61+261512.4)           & Seyfert & 0.30801 & 180.235924 & $+$26.253449 & 14 \\
          & SDSS\,J115552.80+292238.4 (QSO\,J1155+2922)            & Quasar  & 0.51979 & 178.969995 & $+$29.377345 & 10 \\
\hline
NGC\,3992 & SBS 1213+549 (MCG+09-20-133,RX\,J1215.8+5442)          & Seyfert & 0.15007 & 183.956019 & $+$54.706660 & 11 \\
          & SDSS J120506.35+563330.9 (ELARS\,19)                   & Galaxy  & 0.03087 & 181.276488 & $+$56.558585 &  5 \\
          & Mrk\,1447                                              & Seyfert & 0.09558 & 172.621391 & $+$49.582726 & 12 \\
          & SDSS J112448.29+531818.8 ([VV2006]\,J112448.3+531818   & Quasar  & 0.53110 & 171.201218 & $+$53.305237 &  9 \\
          & HS\,1119+5812 (SDSS\,J112244.88+575543.0)              & Quasar  & 0.90572 & 170.687029 & $+$57.928637 & 10 \\
          & SBS\,1116+523 ([VV2006]\,J111948.0+520554)             & Quasar  & 0.35568 & 169.949782 & $+$52.098045 & 16 \\
          & SBS\,1116+518 (RBS\,967,RX\,J1119.6+5133)              & Seyfert & 0.10719 & 169.908418 & $+$51.554311 &  8 \\
          & RX\,J1117.6+5301 (SDSS J111740.48+530151.2)            & Seyfert & 0.15843 & 169.418720 & $+$53.030904 & 12 \\
          & SDSS J111443.65+525834.2 (2MASX\,J11144367+5258338)    & Seyfert & 0.07908 & 168.681909 & $+$52.976190 &  8 \\
\hline
\end{tabular}
\noindent
\\
$^{\rm a}$\,adopted from NED; alternative names are given in parentheses;
$^{\rm b}$\,S/N per resolution element at $\lambda=1230$ \AA.
\end{scriptsize}
\end{table*}

%%%%%%%%%%%%%%%%%%%%%% SECTION 03 %%%%%%%%%%%%%%%%%%%%%%

\section{Discussion of individual absorbers}

\subsection{General absorber statistics}
 
In the final sample of 35 AGN spectra, H\,{\sc i} Ly\,$\alpha$ absorption 
within the groups' velocity range $\Delta v_{\rm abs}$ (gray shaded area in the velocity plots
in Figs.\,2-5) is detected along 19 sightlines (Table 3). Only one of these OGrM absorbers
(towards Ton\,133) possibly has associated metal absorption (Si\,{\sc iii}; see Sect.\,3.4), so that
the vast majority of the systems are considered as ``Ly\,$\alpha$-only absorbers''
\citep[but see also][for a potential group-related population of weak metal absorbers]{muzahid2018}.
In Sects.\,3.2-3.5, we discuss 
the properties of the absorbers individually for each group environment. 

With these numbers, it is already evident that OGrM absorbers are frequent; 
the detection rate comes out to $f_{\rm det} = 19/35 = 0.54$. More meaningful is
the detection rate for a (minimum) limiting H\,{\sc i} column density, $N_{\rm lim}$,
and a pre-defined range in impact parameter.
For log $N_{\rm lim}=13.2$ and $\rho /R_{\rm vir}=1-3$, we have 34 
sightlines with a sufficient S/N to detect H\,{\sc i} Ly\,$\alpha$ absorption at that 
level and 16 OGrM absorbers detected at log $N$(H\,{\sc i}$)\geq 13.2$. This implies
$f_{\rm det,13.2} = 16/34 = 0.47\pm 0.12$ for that impact parameter range,
the large error being a result of the limited absorber statistics. 
In Sect.\,4.2, we place this detection rate in a cosmological context, by comparing the incidence 
rate of Ly\,$\alpha$ absorption in the OGrM  with that of Ly\,$\alpha$ absorption
in the local Ly\,$\alpha$ forest and in nearby filaments. 

The measured Ly\,$\alpha$ equivalent widths in our OGrM absorber sample range 
from $17$ to $505$ m\AA, while the logarithmic H\,{\sc i} column densities 
derived from the component-modeling and AOD methods
have values of log $N$(H\,{\sc i}$)=12.50-14.34$ (see Table 3).
This column-density range is very similar to that of the local Ly\,$\alpha$ forest 
\citep[][see also Sect.\,4.1]{lehner2007,danforth2016,bouma2022}, implying that 
these absorbers are predominantly ionized with only small neutral gas fractions.
The $b$ values derived from the H\,{\sc i} component modeling for the OGrM  absorbers 
lie in the range between $16$ and $80$ km\,s$^{-1}$. Because of the limited
spectral resolution and S/N of the COS data, however, the actual component structure may
not be fully resolved. Therefore, the interpretation of the derived $b$ values 
(in terms of line-broadening mechanisms) remains afflicted with large 
uncertainties (see discussion in Sect.\,4.1).

In Fig.\,2-4, we show velocity plots for the 35 sightlines that pass the OGrM
of the four galaxy groups. In the top panel of each figure, we plot the histograms 
for the radial velocities of the groups' confirmed member galaxies (green) and all 
galaxies found in these fields (gray). The expected velocity interval for 
OGrM Ly\,$\alpha$ absorption is indicated with the gray-shaded range in the 
velocity plots.

%%%%%%%%%%%%%%%%%%%%%%% FIGURE 02 %%%%%%%%%%%%%%%%%%%%%%%

\begin{figure}[t!]
\begin{center}
\resizebox{1.0\hsize}{!}{\includegraphics{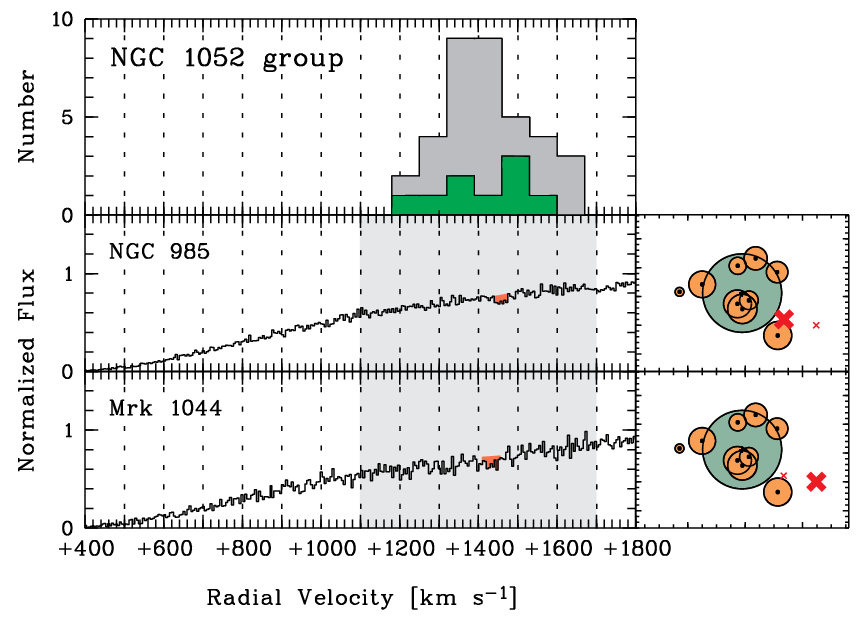}}
\caption[]{
Velocity plots for H\,{\sc i} Ly\,$\alpha$ absorption for the sightlines
passing the outer environment of the NGC\,1052 group (lower two panels).
OGrM absorbers are marked in dark orange. The expected radial velocity range
for OGrM absorption is indicated with the gray-shaded area (see Sect.\,2.3). 
The upper panel shows the velocity distribution for group member galaxies 
(green histogram) and for all galaxies in the group's field (gray histogram).
}
\end{center}
\end{figure}

%%%%%%%%%%%%%%%%%%%%%%% FIGURE 03 %%%%%%%%%%%%%%%%%%%%%%%

\begin{figure}[t!]
\begin{center}
\resizebox{1.0\hsize}{!}{\includegraphics{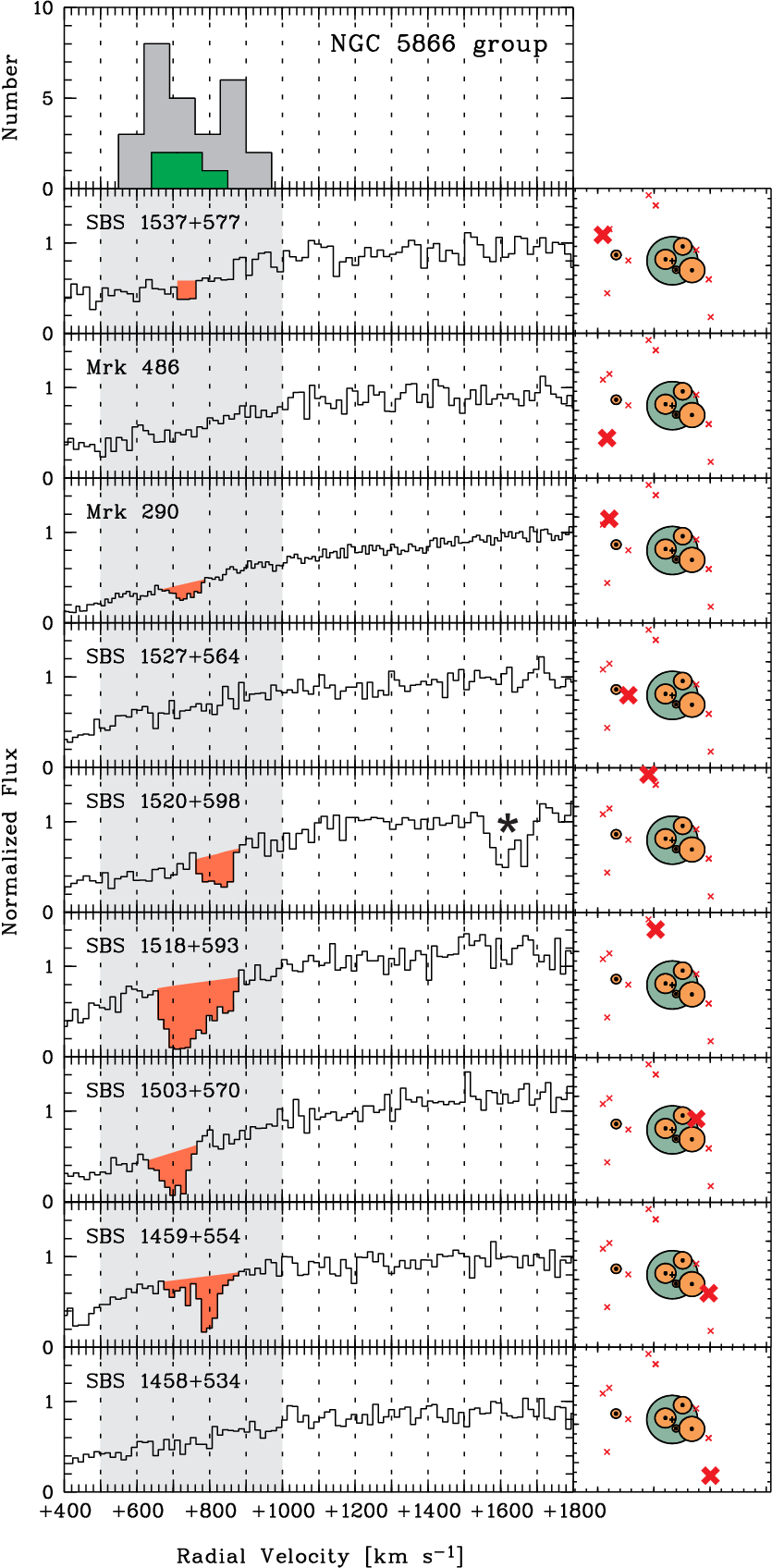}}
\caption[]{
Same as Fig.\,2, but for the NGC\,5866 group. Absorption lines not
related to the OGrM or CGM of member/field galaxies are labeled
with the star symbol. 
}
\end{center}
\end{figure}

%%%%%%%%%%%%%%%%%%%%%%%%%%%%%%%%%%%%%%%%%%%%%%%%%%%%%%

\subsection{NGC\,1052 group}

There are 53 galaxies in the NGC\,1052 group field 
($\alpha_{2000}\approx 44-36$ deg, $\delta_{2000}\approx -11$ to $-5$ deg;
see Fig.\,1) in our galaxy sample with velocities between 1240 and 1680 km\,s$^{-1}$,
from which we regard 9 galaxies as group members.

Two lines of sight (LOS) in our COS sample towards NGC\,985 and Mrk\,1044 pass the 
NGC\,1052 group at normalized impact parameters of $\rho_{\rm GC} /R_{\rm vir}=1.24$ and $2.04$ 
from the group's center (GC) position. H\,{\sc i} Ly\,$\alpha$ absorption 
is detected within the group's radial velocity range along both sightlines 
(see Fig.\,2). Very weak ($<30$ m\AA) but significant
single-component Ly\,$\alpha$ absorption is detected at 1457 km\,s$^{-1}$ (NGC\,985)
and 1433 km\,s$^{-1}$ (Mrk\,1044). Following our definition in Sect.\,2.3,
these absorbers define a coherent absorption pattern
along these neighboring sightlines that are separated by $38$ kpc at
the distance of the NGC\,1052 group. These two sightlines do not pass the 
CGM of any of the galaxies in this field (including group member galaxies
and field galaxies). The S/N per resolution element
in these two spectra is $\geq 50$, thus very high. Because of the weakness of 
these absorption features, we re-analyzed the two original data sets with 
an alternative version of the CALCOS sightline, as recently developed in our 
group for the purpose of analyzing He\,{\sc ii} Ly\,$\alpha$ transmission 
spikes \citep{makan2021}. These features remain, however, clearly present
also in the newly reduced data set with almost no changes in the flux distribution,
leading us to conclude that they represent real intervening absorption features
that trace the OGrM of the NGC\,1052 group.

With column densites of log $N$(H\,{\sc i}$)=12.75$ (Mrk\,1044) and $12.50$ (NGC\,985),
these absorbers represent the weakest systems in our OGrM sample. Note that such weak systems 
would remain unnoticed in COS spectra with S/N$<30$, thus in most archival AGN data
sets.

%%%%%%%%%%%%%%%%%%%%%%% FIGURE 04 %%%%%%%%%%%%%%%%%%%%%%%

\begin{figure*}[t!]
\begin{center}
\resizebox{1.0\hsize}{!}{\includegraphics{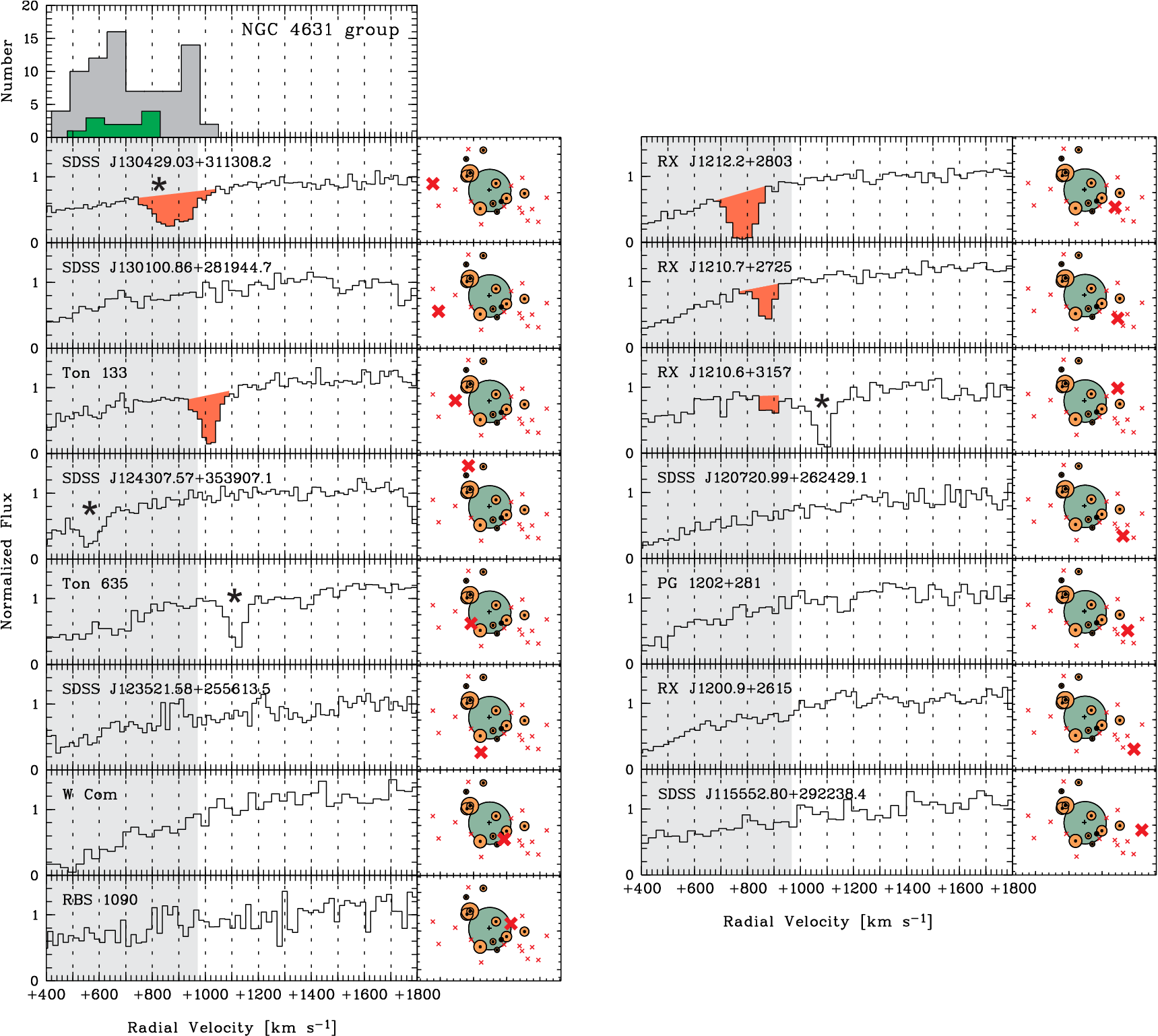}}
\caption[]{
Same as Fig.\,2, but for the NGC\,4631 group. 
}
\end{center}
\end{figure*}

%%%%%%%%%%%%%%%%%%%%%% TABLE 03 %%%%%%%%%%%%%%%%%%%%%%

\begin{table*}[t!]
\caption[]{Summary of absorption-line measurements}
\begin{footnotesize}
\begin{tabular}{lrlrrrrrrr}
\hline
Group name       & LOS\,No. & Sightline           & $v_{\rm abs}$  & $W_{{\rm Ly}\,\alpha}$ & log $N$(H\,{\sc i}) 
                 & $b$(H\,{\sc i}) & $\rho_{\rm GC}/R_{\rm vir}$$^{\rm a}$  & $d_N$$^{\rm b}$ \\
                 &              &                 & [km\,s$^{-1}$] & [m\AA]                 
                 &              & [km\,s$^{-1}$]  &                & [kpc] \\
\hline
\hline
NGC\,1052        &  1 & NGC\,985                  &  1457          & $17\pm5$           & $12.50\pm0.09$      & $19\pm5$        & $1.24$  &  38     \\
                 &  2 & Mrk\,1044                 &  1433          & $27\pm8$           & $12.75\pm0.14$      & $32\pm12$       & $2.04$  &  38     \\

\hline
NGC 5866         &  3 & SBS\,1537+577             &   735          & $59\pm14$          & $13.06\pm0.15$      & $16\pm6$        & $2.89$  &  12     \\
                 &  4 & Mrk\,486                  &   ...          & $\leq 48$          & $\leq 12.95$        & ...             & $2.94$  &  ...    \\
                 &  5 & Mrk\,290                  &   735          & $129\pm10$         & $13.47\pm0.04$      & $42\pm4$        & $2.75$  &  12     \\
                 &  6 & SBS\,1527+564             &   ...          & $\leq 50$          & $\leq 12.96$        & ...             & $1.73$  &  ...    \\
                 &  7 & SBS\,1520+598             &   831          & $168\pm22$         & $13.63\pm0.07$      & $44\pm7$        & $2.86$  &  17     \\
                 &  8 & SBS\,1518+593             &   725          & $505\pm24$         & $14.16\pm0.05$      & $43\pm6$        & $2.38$  &  17     \\
                 &    &                           &   826          &                    & $13.49\pm0.08$      & $43\pm7$        &         &  ..     \\
                 &  9 & SBS\,1503+570             &   703          & $284\pm25$         & $13.96\pm0.15$      & $35\pm8$        & $1.04$  &  46     \\
                 & 10 & SBS\,1459+554             &   721          & $228\pm28$         & $13.07\pm0.13$      & $37\pm13$       & $1.64$  &  46     \\
                 &    &                           &   791          &                    & $13.71\pm0.06$      & $22\pm4$        &         &  ...    \\
                 & 11 & SBS\,1458+534             &   ...          & $\leq 51$          & $\leq 12.97$        & ...             & $2.81$  &  ...    \\
\hline
NGC\,4631        & 12 & SDSS\,J130429.03+311308.2 &   923          & $433\pm34$         & $13.85\pm0.11$      & $53\pm11$       & $2.67$  &  47     \\ 
                 & 13 & SDSS\,J130100.86+281944.7 &   ...          & $\leq 46$          & $\leq 12.93$        & ...             & $2.53$  &  ...    \\ 
                 & 14 & Ton\,133                  &  1013          & $272\pm22$         & $13.92\pm0.05$      & $35\pm05$       & $1.61$  &  47     \\    
                 & 15 & SDSS\,J124307.57+353907.1 &   ...          & $\leq 28$          & $\leq 12.71$        & ...             & $2.15$  &  ...    \\        
                 & 16 & Ton\,635                  &   ...          & $\leq 46$          & $\leq 12.93$        & ...             & $1.02$  &  ...    \\     
                 & 17 & SDSS\,J123521.58+255613.5 &   ...          & $\leq 53$          & $\leq 12.99$        & ...             & $1.66$  &  ...    \\    
                 & 18 & WCom                      &   ...          & $\leq 85$          & $\leq 13.19$        & ...             & $1.06$  &  ...    \\       
                 & 19 & RBS\,1090                 &   ...          & $\leq 102$         & $\leq 13.27$        & ...             & $1.06$  &  ...    \\       
                 & 20 & RX\_J1212.2+2803          &   788          & $425\pm20$         & $14.34\pm0.08$      & $40\pm4$        & $1.68$  &  11     \\
                 & 21 & RX\,J1210.7+2725          &   883          & $151\pm15$         & $13.58\pm0.05$      & $25\pm4$        & $1.91$  &  11     \\  
                 & 22 & RX\,J1210.6+3157          &   895          & $85\pm27$          & $13.45\pm0.22$      & $36\pm9$        & $1.66$  &  61     \\     
                 & 23 & SDSS\,J120720.99+262429.1 &   ...          & $\leq 49$          & $\leq 12.95$        & ...             & $2.36$  &  ...    \\      
                 & 24 & PG\,1202+281              &   ...          & $\leq 51$          & $\leq 12.97$        & ...             & $2.25$  &  ...    \\      
                 & 25 & RX\,J1200.9+2615          &   ...          & $\leq 39$          & $\leq 12.86$        & ...             & $2.81$  &  ...    \\    
                 & 26 & SDSS\,J115552.80+292238.4 &   ...          & $\leq 49$          & $\leq 12.95$        & ...             & $2.77$  &  ...    \\      
\hline
NGC 3992         & 27 & SBS\,1213+549             &   ...          & $\leq 39$          & $\leq 12.86$        & ...             & $1.92$  &  ...    \\
                 & 28 & SDSS\,J120506.35+563330.9 &   ...          & $\leq 79$          & $\leq 13.16$        & ...             & $1.77$  &  ...    \\      
                 & 29 & Mrk\,1447                 &   772          & $342\pm22$         & $14.14\pm0.08$      & $37\pm4$        & $2.44$  &  92     \\
                 & 30 & SDSS\,J112448.29+531818.8 &   658          & $285\pm42$         & $13.86\pm0.05$      & $43\pm5$        & $1.80$  &  38     \\
                 & 31 & HS\,1119+5812             &   ...          & $\leq 60$          & $\leq 13.04$        & ...             & $2.77$  &  ...    \\
                 & 32 & SBS\,1116+523             &   723          & $250\pm18$         & $13.86\pm0.04$      & $40\pm4$        & $2.31$  &  19     \\           
                 & 33 & SBS\,1116+518             &   715          & $220\pm59$         & $13.68\pm0.08$      & $53\pm12$       & $2.43$  &  19     \\
                 & 34 & RX\,J1117.6+5301          &   676          & $268\pm31$         & $13.87\pm0.05$      & $37\pm5$        & $2.55$  &  16     \\
                 & 35 & SDSS\,J111443.65+525834.2 &   675          & $131\pm36$         & $13.53\pm0.11$      & $80\pm23$       & $2.34$  &  16     \\
\hline
\end{tabular}
\noindent
\\
$^{\rm a}$\,$R_{\rm vir}$-normalized impact parameter to group center.\\
$^{\rm b}$\,Distance to neighboring OGrM sightline.\\
\end{footnotesize}
\end{table*}

%%%%%%%%%%%%%%%%%%%%%%%%%%%%%%%%%%%%%%%%%%%%%%%%%%%%%%

\subsection{NGC\,5866 group}

The NGC\,5866 group field is relatively sparse with only 27 galaxies in the
range $\alpha_{2000}\approx 237-220$ deg, $\delta_{2000}\approx 53-60$ deg at velocities
$v=560-960$ km\,s$^{-1}$, from which 5 galaxies are regarded as secure
group members. The group's velocity dispersion ($\sigma_{\rm gr}=82$ km\,s$^{-1}$)
and total luminosity are small, resulting in a rather small group mass and virial
radius (see Table 1).

The large-scale environment of the NGC\,5866 group at $\rho_{\rm GC} /R_{\rm vir}=1-3$ 
is sampled by nine AGN sightlines that are not contaminated by CGM absorption (see Sect.\,2.3).
The normalized impact parameters range from $1.04$ to $2.94$.
The COS spectral data are of partly good, partly mediocre quality with a S/N for most spectra 
of  $\leq 15$ (see Table 2). The majority of the sightlines (6/9) are located 
in the eastern region of the group (Fig.\,1).
Significant OGrM Ly\,$\alpha$ absorption in this velocity
range is detected along six sightlines, forming a coherent OGrM absorption pattern (see definition
in Sect.\,2.3 and velocity plots in Fig.\,3). 
The absorption velocities scatter around $\sim 800$ km\,s$^{-1}$.

The two relatively strong and broad absorbers towards SBS\,1518+593 and SBS\,1459+554
have an asymmetrical shape, which can be modeled by a two-component structure (Table 3). However,
no associated metal absorption is detected in these systems, same as for the other seven
OGrM absorbers in the NGC\,5866 group environment.
The derived H\,{\sc i} column densities for OGrM systems around the NGC\,5866 group
are between log $N$(H\,{\sc i}$)=13.06$ and $14.16$ (see Table 3), thus
substantially higher than for the NGC\,1052 group. 

We note that the NGC\,5866 group environment and the Ly\,$\alpha$ absorbers discussed 
here were also included in the study of \citet{wakker2015} and identified as filament 
absorbers. This is no contradiction to our results, as OGrM absorbers (as defined
in Sect.\,1) represent a sub-class of filament absorbers that are spatially and 
kinematically coupled to knots in the Cosmic Web (i.e., to groups).

%%%%%%%%%%%%%%%%%%%%%%% FIGURE 05 %%%%%%%%%%%%%%%%%%%%%%%

\begin{figure}[t!]
\begin{center}
\resizebox{1.0\hsize}{!}{\includegraphics{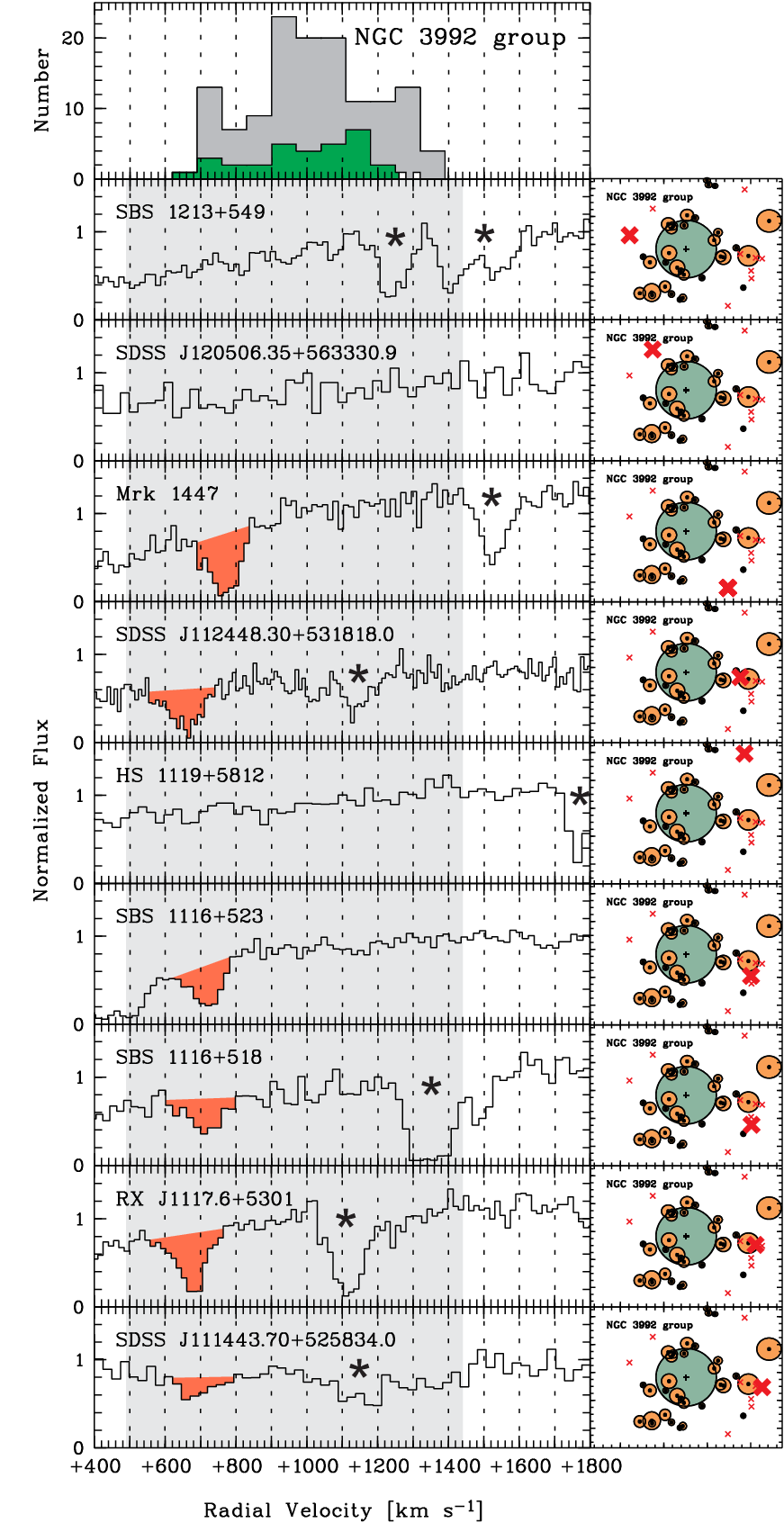}}
\caption[]{
Same as Fig.\,2, but for the NGC\,3992 group.
}
\end{center}
\end{figure}

%%%%%%%%%%%%%%%%%%%%%%%%%%%%%%%%%%%%%%%%%%%%%%%%%%%%%%%%%

\subsection{NGC\,4631 group}

The NGC\,4631 group field ($\alpha_{2000}\approx 198-177$ deg, $\delta_{2000}\approx 24-37$ deg) 
is densley populated with 114 galaxies in the velocity range between 430 and 1030 km\,s$^{-1}$.
From these, however, only 12 galaxies are regarded to be secure members of the NGC\,4631 group
\citep{fouque1992}.

There are 15 sightlines in our AGN sample that pass the outer region of the 
NGC\,4631 group beyond its virial radius at normalized impact parameters between 
$\rho_{\rm GC}/R_{\rm vir}=1.02$ and $2.77$ and at medium to low S/N (see Tables 2 and 3). 
Six of these sightlines sample the eastern part of the group at right ascensions $>192$ deg, 
while the remaining nine sightlines cluster in the south-western region of the group.

Only along five out of the 15 sightlines is OGrM Ly\,$\alpha$ 
absorption detected, implying an OGrM detection rate that is substantially smaller
than in the other three groups. The absorption velocities range between $\sim 700$ and 
$\sim 1100$ km\,s$^{-1}$, while the equivalent widths have values between 
$85$ and $433$ m\AA\,(see Table 3 and velocity plots in Fig.\,4).

The OGrM absorber towards Ton\,133 is the only OGrM system in our sample that potentially 
shows associated metal absorption (here: Si\,{\sc iii} $\lambda 1206.5$). Since no other 
metal lines are detected, however, the identification of the Si\,{\sc iii} absorption 
remains ambiguous. The very broad ($b=53$ km\,s$^{-1}$) OGrM absorber towards the neighboring 
sightline SDSS\,J130429.03+311308.2 is potentially blended in the blue wing with intervening 
C\,{\sc iii} $\lambda 977.0$ absorption at $z_{\rm abs}=0.2477$.

Following our definition in Sect.\,2.3, these two absorbers form a coherent OGrM absorption 
pattern in the eastern part of the group, while the three absorbers along the neighboring
sightlines towards RX\,J1212.2+2803, RX\,J1210.7+2725, and RX\,J1210.6+3157
form another coherent OGrM pattern in the south-western part of the NGC\,4631 group outskirts.

Note that the immediate environment of NGC\,4631 itself is characterized by a gigantic tidal gas 
stream that stems from the interaction of NGC\,4631 with its neighboring galaxies
\citep[see][their Fig.\,2]{richter2018}.

\subsection{NGC\,3992 group}

The relatively large and well-populated NGC\,3992 group field ($\alpha_{2000}\approx 188-167$ deg, 
$\delta_{2000}\approx 49-59$ deg) contains 236 galaxies at $v=590-1430$ km\,s$^{-1}$, 
32 of these are regarded as members of the NGC\,3992 group.

Seven of the nine selected sightlines for this group pass the western outer environment of the 
NGC\,3992 group at impact parameters between $\rho_{\rm GC}/R_{\rm vir}=1.77$ and $2.77$ (see Table 3).
They sample an outer region of the group that still contains several massive galaxies (see Fig.\,1).
Six of these seven sightlines exhibit moderate to strong OGrM Ly\,$\alpha$ absorption in a coherent
absorption pattern near $700$ km\,s$^{-1}$ with Ly\,$\alpha$ equivalent widths between $131$ 
and $342$ m\AA\, and column densities in the range log $N$(H\,{\sc i}$)=13.53-14.14$.

The two sightlines towards SBS\,1213+549 and SDSS\,J120506.35+563330.9 in the eastern region at
$\alpha_{2000}>180$ deg do not show any significant Ly\,$\alpha$ absorption in the velocity 
range of the NGC\,3992 group.

%%%%%%%%%%%%%%%%%%%%%% SECTION 04 %%%%%%%%%%%%%%%%%%%%%%

\section{Nature and origin of the OGrM absorbers}

\subsection{Observational trends}

With a detection rate of $\sim 50$ percent in our survey, kinematically aligned 
OGrM Ly\,$\alpha$ absorbers appear to represent a typical absorber class 
in the outer regions of nearby galaxy groups, where these connect with the 
ambient Cosmic Web. Per definition, OGrM Ly\,$\alpha$ absorbers are kinematically 
aligned with their host groups to ensure that they trace the same overall cosmological 
environment. But what does this alignment tell us about the nature of these systems?

On the one hand, the coherence in velocity along multiple sightlines could indicate
that OGrM absorbers trace the same physical structure (i.e., one and the same 
``cloud'') probed at different locations, that (for instance) is floating
from the Cosmic Web towards the group center (scenario I). On the other hand, the 
absorbers may represent independent gas structures (tracing density peaks)
that move (``swarm-like'') as an ensemble of clouds within the group's outer
potential well (scenario II). 
Depending on the actual physical size and the separation between the sightlines
at the distances of the groups, one could also think of a combination of 
both of these scenarios.

To visualize the geometrical setup, we have plotted in Fig.\,6 the 
positions of the OGrM absorbers with respect to the confirmed group member galaxies 
in 2D and 3D projections (with the velocity as third dimension). The dashed
line indicates for each LOS the nearest neighboring sightline that also hosts an OGrM absorber. 
The linear distance to the nearest neighboring OGrM sightline, $d_N$, as calculated from
the angular separation of the LOS and the group distance, is listed for 
each OGrM absorber in the last column of Table 3. 
As can be seen, values for $d_N$ range from 11 to 92 kpc, a linear scale in which
both scenarios mentioned above appear realistic. We will elaborate further on the
absorber sizes in Sect.\,4.4.

In Fig.\,7, we plot the measured Ly\,$\alpha$ equivalent widths and the derived 
H\,{\sc i} column densities against the normalized absorber impact parameter to the 
group center. For the entire impact parameter range ($\rho /R_{\rm vir}=1-3$), there 
is a lot of scatter with no obvious trend. In future studies of the OGrM, using a larger 
absorber sample, it would be interesting to also systematically explore the range 
$\rho /R_{\rm vir}>3$ to search for observational trends that may (or may not) indicate  
a change in the physical conditions of the absorbers related
to the decreasing cosmological matter overdensity in this interface region between
collapsed group structures and the ambient Cosmic Web.

The overall distribution of H\,{\sc i} column densities (lower right panel in Fig.\,7) 
and the median value for log $N$(H\,{\sc i}) of $13.68$ are very similar to those derived 
for low-redshift IGM filaments \citep{bouma2022,wakker2015} or in ``blind'' Ly\,$\alpha$-forest 
surveys \citep{lehner2007,danforth2016}, implying that the average physical 
conditions are comparable. This is not really surprising, however, as OGrM absorbers
represent a sub-class of the overall intergalactic Ly\,$\alpha$ population
and thus are included in the forementioned surveys.

To explore in more detail the origin and physical conditions of the OGrM systems, one would 
like to know about the ionization conditions and gas densities in the absorbers, from which one
also could derive a characteristic size. With H\,{\sc i} Ly\,$\alpha$ absorption as 
the only information carrier for all but one of the absorbers, however, we have only a very 
limited diagnostic toolkit at hand to explore the physical conditions in the gas.
Note that the lack of metal-line absorption in our sample of OGrM systems does not 
necessarily point toward a low metallicity of the OGrM, but rather is a result of the 
low gas column- and volume densities and the high degree of ionization, which is 
typical for intergalactic gas at such small cosmological overdensities
\citep[e.g.][see also Sect.\,4.3]{martizzi2019}.

The directly measured quantities available for all OGrM absorbers in our sample are: H\,{\sc i} 
column density, Doppler parameter, radial velocity, and the component structure. From these parameters
alone, we can derive estimates for the gas densities, total gas column densities, and absorber sizes only using 
certain assumptions (Sect.\,4.3).

In principle, the measured $b$ value in each absorber provides an upper 
temperature limit for the gas, as the thermal component of the Doppler parameter can be 
written as $b_{\rm th}^2=2kT/m_{\rm H}$, with $m_{\rm H}$ being the mass of the 
hydrogen atom. It is, however, commonly assumed that the observed $b$ value is composed of a
thermal and a non-thermal component, so that $b^2=b_{\rm th}^{\,2}+b_{\rm non-th}^{\,2}$.
In our sample, the limited S/N and relatively low spectral resolution complicates 
the interpretation of the measured H\,{\sc i} $b$ values as temperature indicator because the 
component sub-structure may not be fully resolved. The measured $b$ values in the 
OGrM absorber-components are all $\leq 53$ km\,s$^{-1}$ (see Table 3), with only one 
exception (the weak OGrM absorber towards SDSS\,J111443.65+525834.2, for which $b$ 
is only poorly constrained; see last row in Table 3). A value of $b=53$ km\,s$^{-1}$
translates into an upper
temperature limit of $T\leq 1.7 \times 10^5$ K. This value is an order of magnitude lower 
than the expected virial temperatures of the groups ($T>4 \times 10^6$ K) and lower
than what is expected for (equilibrium) collisional ionization to be dominant
\citep[$T>2 \times 10^5$ K;][]{richter2008}.
While we cannot exclude that non-equilibrium collisional ionization \citep[e.g.,][]{gnat2007}
and/or a mixture of collisional and photo-ionization are relevant,
the moderate $b$ values for the majority of the systems suggest that the OGrM absorbers are predominantly 
photoionized (possibly being embedded in a hotter, collisionally ionized gas phase).

%%%%%%%%%%%%%%%%%%%%%%% FIGURE 06 %%%%%%%%%%%%%%%%%%%%%%%

\begin{figure*}[t!]
\sidecaption
\resizebox{0.70\hsize}{!}{\includegraphics{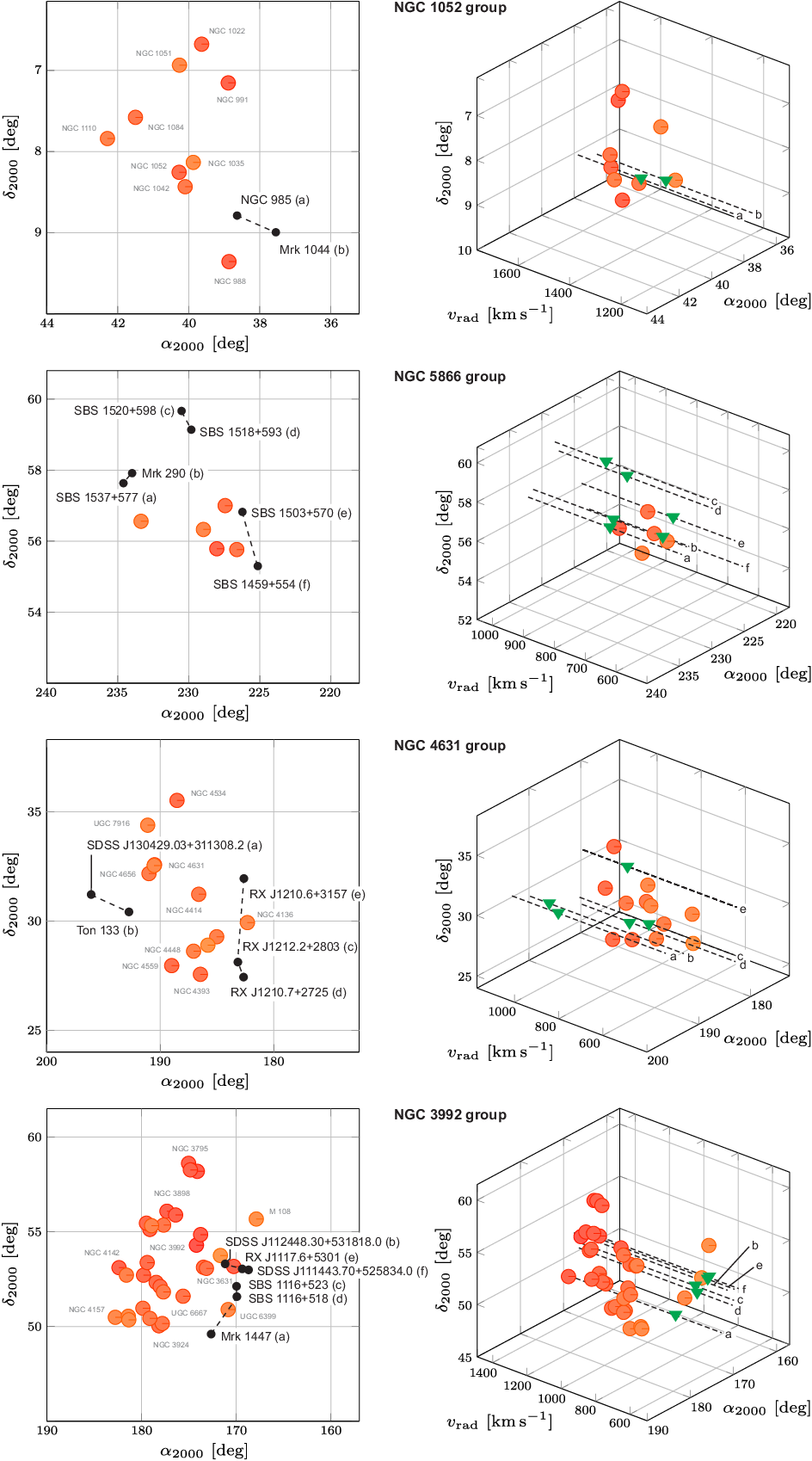}}
\caption[]{
Spatial configuration of the group member galaxies and the AGN sightlines for the
four groups in our sample together with supplementary information.
The left panel shows the projected position of group-member galaxies (orange dots)
and the AGN sightlines along which OGrM absorption has been detected (black dots).
The AGN names are labeled in black, the most important group member galaxies are
labeled in gray.
The right panel shows the group-member galaxies (orange dots) and the OGrM systems
(green triangles) in a 3D position-velocity diagram.
}
\end{figure*}

%%%%%%%%%%%%%%%%%%%%%%%%%%%%%%%%%%%%%%%%%%%%%%%%%%%%%%%%%

\subsection{On the incidence rate of OGrM absorbers}

The Ly\,$\alpha$ line density in an AGN spectrum usually is represented by 
$d{\cal N}/dz$, the number of Ly\,$\alpha$ absorbers per unit redshift, which 
characterizes the frequency and distribution of gas clouds in the Cosmic Web 
along that line of sight. The values for $d{\cal N}/dz$ in ``blind'' $z\approx 0$ 
Ly\,$\alpha$ surveys, where $dz$ is calculated from the redshift paths for multiple AGN 
sightlines, range from $d{\cal N}/dz\approx 80$ to $100$ for 
log $N$(H\,{\sc i}$)\geq 13.2$ \citep{danforth2016,lehner2007}. Because 
$v=cz$ at $z\approx 0$, a value of $d{\cal N}/dz=100$ for log $N$(H\,{\sc i}$)\geq 13.2$ 
implies that one expects - on average - one Ly\,$\alpha$ absorber 
per $3000$ km\,s$^{-1}$ wide velocity interval (above that column-density level).

Ly\,$\alpha$ lines are not randomly distributed in redshift space, however, as
they follow the small- and large-scale matter distribution 
(i.e., galaxies, filaments, groups, clusters) in the Cosmic Web \citep[e.g.][]{morris1993}. 
The positions of Ly\,$\alpha$ lines along a AGN spectrum thus are determined 
by both the cosmological expansion and the peculiar gas motions within collapsed stuctures. 
Strong absorbers with log $N$(H\,{\sc i}$)\geq 15$ appear to be highly correlated
with individual galaxies \citep[e.g.][]{prochaska2011}, while weaker systems 
tend to partly cluster in filaments \citep{bouma2022} or partly are part of a
population of ``random'' Ly\,$\alpha$ absorbers that reside in under-dense regions
\citep{tejos2012}. \citet{bouma2022} derived for their sample of velocity-selected 
Ly\,$\alpha$ absorbers within $z=0$ filaments an incidence rate of $d{\cal N}/dz=189\pm 25$
for log $N$(H\,{\sc i}$)\geq 13.2$. This is substantially higher than for the
overall (unselected) Ly\,$\alpha$ population, reflecting the clustering of 
Ly\,$\alpha$ absorbers in cosmological filaments.

To compare the incidence rate of OGrM absorbers with that of filament absorbers 
and the overall Ly\,$\alpha$ forest, we follow the same approach as in \citet{bouma2022} 
and transform the observed OGrM detection rate into a $d{\cal N}/dz$ value.
The total OGrM absorption path length in our sample (in velocity units) is given by 

\begin{equation}
\Delta v_{\rm OGrM}= \sum_{i=1}^4\,{\cal N}_{13.2,i}\,f_i\,\Delta v_i,
\end{equation}

where $i$ is the index for our 4 groups, ${\cal N}_{13.2,i}$ is the number of sightlines
per group that have sufficient S/N to detect OGrM Ly\,$\alpha$ absorption at 
log $N$(H\,{\sc i}$)\geq 13.2$, $f_i\leq1$ is a factor that corrects for line blends,
and $\Delta v_i$ is the expected velocity range for OGrM absorption in group $i$,
as defined in Sect.\,2.3. Since we have rigorously sorted out sightlines that
are contaminated by CGM absorption within $\Delta v_i$, we can safely set $f_i =1$.

Based on equation (4), the total OGrM velocity path in our sample comes out to
$\Delta v_{\rm OGrM}=20,650$ km\,s$^{-1}$. Because $v=cz$ at $z=0$, this velocity
range corresponds to an OGrM redshift pathlength of $dz=\Delta v_{\rm OGrM}/c=0.06888$.
Note that the OGrM redshift pathlength calculated in this way has no cosmological meaning; it just indicates
the integrated velocity range for OGrM absorption in our AGN sample in unit redshift ($\Delta v/c$).
Having 16 OGrM absorbers at $\rho /R_{\rm vir}=1-3$ with log $N$(H\,{\sc i}$)\geq 13.2$, we thus derive 
$d{\cal N}/dz=16/0.06888=232\pm 58$. This is more than twice the number density in the 
overall Ly\,$\alpha$ forest ($d{\cal N}/dz=80-100$) and $\sim 25$ percent above the 
value for $z=0$ filaments ($189$; see above).

Our study therefore indicates that the outer regions of groups at $\rho /R_{\rm vir}=1-3$ are characterized 
by an overdensity of Ly\,$\alpha$ absorbers when compared to other regions in the local Cosmic Web.
The high value for $d{\cal N}/dz$ in the OGrM needs to be, however, re-evaluated and confirmed by a larger 
group/absorber sample.

%%%%%%%%%%%%%%%%%%%%%%% FIGURE 07 %%%%%%%%%%%%%%%%%%%%%%%

\begin{figure}[h!]
\begin{center}
\resizebox{0.95\hsize}{!}{\includegraphics{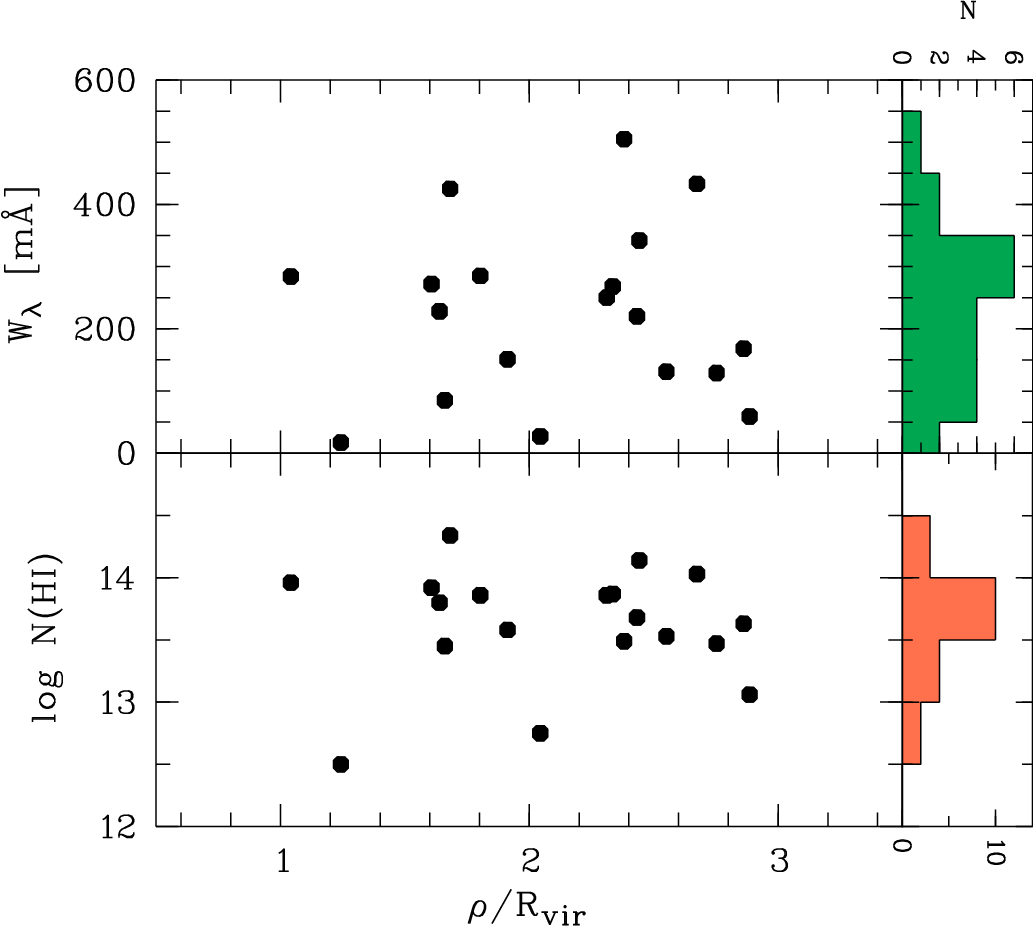}}
\caption[]{
The Ly\,$\alpha$ equivalent widths and logarithmic H\,{\sc i} column densities
of OGrM absorbers are plotted against the relative absorber impact parameter to
the group center. No clear trend can be seen in these plots.
}
\end{center}
\end{figure}

%%%%%%%%%%%%%%%%%%%%%%%%%%%%%%%%%%%%%%%%%%%%%%%%%%%%%%%%%

\subsection{Hydrostatic model}

To shed more light into the physical conditions of the OGrM absorber population, we discuss a simple
hydrostatic toy model, in which we regard the absorbers as mildly overdense gas patches that have
neutral gas fractions high enough to be detectable in H\,{\sc i} Ly\,$\alpha$ absorption
and that are photoionized by the cosmic UV background. The fundamental assumption in this 
model is that the absorbing structures stay roughly in pressure equilibrium with their
ambient gaseous environment and that the gas is in hydrostatic equilibrium, gravitationally 
confined by the underlying matter distribution.

We are aware of the fact that such a model represents a rather over-idealized approach,
as the assumption of hydrostatic equilibrium may not hold for gas in the outskirts of 
low-mass galaxy groups \citep{paul2017}. However, in the absence of any alternative, 
more reliable approach, we are interested to explore the outcome of such a model 
to get a general idea of realistic absorber densities and sizes that we then can 
relate to the observed OGrM absorption pattern and the angular separation between 
the sightlines.

We start our model description with the DM mass distribution within $3R_{\rm vir}$ of the 
group centers, which we assume to follow a NFW profile (see also Sect.\,2.2). Using the formalism 
described in \citet{richter2020}, we calculate the expected radial gas pressure profiles of the gas bound 
to the groups within $3R_{\rm vir}$ as a function of the groups' total masses, 
$M_{L}$ (see Table 1). If we further assume that the projected distances of the absorbers 
to the group centers (i.e., the impact parameters) are equal to the actual radial distances, 
$R$, we can obtain in this way an estimate for $P/k$ for each absorber. The resulting values 
are listed in Table 4, fourth row. With the thermal pressure being $P/k=n_{\rm H}T$,
ignoring non-thermal pressure components 
as well as assuming an upper temperature limit of log $T=4.5$
for photoionized gas, we then obtain a lower limit for $n_{\rm H}$ for each absorber
from this model. The results for $n_{\rm H}$ are shown in Table 4, sixth row. The values 
that we obtain from this approach vary from log $n_{\rm H}=-5.00$ to $-3.72$, a density range
that is characteristic of intergalactic/intra-group environments and the Ly\,$\alpha$ forest,
as seen in hydrodynamical cosmological simulations \citep{martizzi2019}. A direct consequence of
our hydrostatic group model with constant $T$ is that $n_{\rm H}$ (such as $P/k$) decreases 
quickly with increasing distance $R$ from the group center with a trend depending on
the total group mass \citep[][their equation A.12]{richter2020}. This trend is shown for all four
groups in Fig.\,B.1 in the Appendix. 

%%%%%%%%%%%%%%%%%%%%%% TABLE 04 %%%%%%%%%%%%%%%%%%%%%%

\begin{table*}[t!]
\caption[]{Modeled physical conditions in the OGrM absorbers}
\begin{footnotesize}
\begin{tabular}{lrlrrrrrrr}
\hline
Group name       & LOS & Sightline   &  log $P/k$  &  log $T$       &  log $n_{\rm H}$  &  log $N$(H)   &  log $f_{\rm HI}$ &  log $L$ & $L/d_N$   \\
                 & No. &             &             &                &                     &               &                   &                      \\
\hline
\hline
NGC\,1052        &  1 & NGC\,985                  &  $1.12$  &  $\leq 4.5$ &  $\geq -3.72$ &  $\leq 15.86$ &  $\geq -3.36$ &  $\leq 1.09$ & $\leq 3106.9$ \\
                 &  2 & Mrk\,1044                 &  $0.50$  &  $\leq 4.5$ &  $\geq -4.34$ &  $\leq 16.72$ &  $\geq -3.97$ &  $\leq 2.57$ & $\leq 102.7$ \\
\hline
NGC 5866         &  3 & SBS\,1537+577             & $-0.16$  &  $\leq 4.5$ &  $\geq -5.00$ &  $\leq 17.69$ &  $\geq -4.63$ &  $\leq 4.20$ & $\leq 0.8$ \\
                 &  5 & Mrk\,290                  & $-0.11$  &  $\leq 4.5$ &  $\geq -4.95$ &  $\leq 18.05$ &  $\geq -4.58$ &  $\leq 4.51$ & $\leq 0.4$ \\
                 &  7 & SBS\,1520+598             & $-0.16$  &  $\leq 4.5$ &  $\geq -5.00$ &  $\leq 18.26$ &  $\geq -4.63$ &  $\leq 4.77$ & $\leq 0.3$ \\
                 &  8 & SBS\,1518+593             &  $0.08$  &  $\leq 4.5$ &  $\geq -4.76$ &  $\leq 18.56$ &  $\geq -4.40$ &  $\leq 4.83$ & $\leq 0.3$ \\
                 &    &                           &  $0.08$  &  $\leq 4.5$ &  $\geq -4.76$ &  $\leq 17.89$ &  $\geq -4.40$ &  $\leq 4.16$ & $\leq 1.2$ \\
                 &  9 & SBS\,1503+570             &  $1.10$  &  $\leq 4.5$ &  $\geq -3.74$ &  $\leq 17.34$ &  $\geq -3.38$ &  $\leq 2.59$ & $\leq 118.0$ \\
                 & 10 & SBS\,1459+554             &  $0.54$  &  $\leq 4.5$ &  $\geq -4.30$ &  $\leq 17.00$ &  $\geq -3.93$ &  $\leq 2.81$ & $\leq 71.1$ \\
                 &    &                           &  $0.54$  &  $\leq 4.5$ &  $\geq -4.30$ &  $\leq 17.64$ &  $\geq -3.93$ &  $\leq 3.45$ & $\leq 16.3$ \\
\hline
NGC\,4631        & 12 & SDSS\,J130429.03+311308.2 &  $0.12$  &  $\leq 4.5$ &  $\geq -4.72$ &  $\leq 17.93$ &  $\geq -4.36$ &  $\leq 4.16$ & $\leq 1.7$ \\
                 &    &                           &  $0.12$  &  $\leq 4.5$ &  $\geq -4.72$ &  $\leq 18.21$ &  $\geq -4.36$ &  $\leq 4.44$ & $\leq 3.2$ \\
                 & 14 & Ton\,133                  &  $0.76$  &  $\leq 4.5$ &  $\geq -4.08$ &  $\leq 17.64$ &  $\geq -3.72$ &  $\leq 3.23$ & $\leq 27.6$ \\
                 & 20 & RX\_J1212.2+2803          &  $0.70$  &  $\leq 4.5$ &  $\geq -4.14$ &  $\leq 18.11$ &  $\geq -3.77$ &  $\leq 3.76$ & $\leq 1.9$ \\
                 & 21 & RX\,J1210.7+2725          &  $0.54$  &  $\leq 4.5$ &  $\geq -4.30$ &  $\leq 17.51$ &  $\geq -3.93$ &  $\leq 3.32$ & $\leq 5.2$ \\
                 & 22 & RX\,J1210.6+3157          &  $0.72$  &  $\leq 4.5$ &  $\geq -4.12$ &  $\leq 17.20$ &  $\geq -3.75$ &  $\leq 2.83$ & $\leq 89.7$ \\
\hline
NGC 3992         & 29 & Mrk\,1447                 &  $0.68$  &  $\leq 4.5$ &  $\geq -4.16$ &  $\leq 17.93$ &  $\geq -3.79$ &  $\leq 3.60$ & $\leq 23.3$ \\
                 & 30 & SDSS\,J112448.29+531818.8 &  $1.06$  &  $\leq 4.5$ &  $\geq -3.78$ &  $\leq 17.27$ &  $\geq -3.41$ &  $\leq 2.56$ & $\leq 104.3$ \\
                 & 32 & SBS\,1116+523             &  $0.75$  &  $\leq 4.5$ &  $\geq -4.09$ &  $\leq 17.58$ &  $\geq -3.72$ &  $\leq 3.18$ & $\leq 12.4$ \\
                 & 33 & SBS\,1116+518             &  $0.69$  &  $\leq 4.5$ &  $\geq -4.15$ &  $\leq 17.46$ &  $\geq -3.78$ &  $\leq 3.12$ & $\leq 14.3$\\
                 & 34 & RX\,J1117.6+5301          &  $0.74$  &  $\leq 4.5$ &  $\geq -4.10$ &  $\leq 17.60$ &  $\geq -3.73$ &  $\leq 3.21$ & $\leq 9.9$ \\
                 & 35 & SDSS\,J111443.65+525834.2 &  $0.63$  &  $\leq 4.5$ &  $\geq -4.21$ &  $\leq 17.37$ &  $\geq -3.85$ &  $\leq 3.10$ & $\leq 12.8$ \\
\hline
\end{tabular}
\noindent
\\
Parameters and their units: gas pressure $P/k$ in [cm$^{-3}$\,K]; gas temperature $T$ in [K],
hydrogen volume density $n_{\rm H}$ in [cm$^{-3}$],\\
total hydrogen column density $N$(H) in [cm$^{-2}$], neutral hydrogen fraction $f_{\rm HI}$,
absorption path length $L$ in [pc]. \\
\end{footnotesize}
\end{table*}

%%%%%%%%%%%%%%%%%%%%%%%%%%%%%%%%%%%%%%%%%%%%%%%%%%%%%%

In our model, we regard photoionization by the cosmic UV background as the only relevant 
ionization process to ionize hydrogen in the OGrM absorbers. This is motivated by the fact that the 
OGrM absorbers are far away from any local ionization sources (such as AGN or stars) and is
supported by the temperature limits from the measured $b$ values (see above). We adopt a 
photoionization rate of $\Gamma = 7 \times 10^{-14}$ s$^{-1}$ for the $z=0$ IGM, based on
the results from \citet[][]{fumagalli2017}.
In analogy to H\,{\sc i} absorbers in the $z=0$ Ly\,$\alpha$ forest, we then can estimate the neutral
gas fraction, $f_{\rm HI}$, in the absorbers assuming photoionization equilibrium and using the
simple relation 

\begin{equation}
f_{\rm HI}= 5.3\,n_{\rm H}T_4^{-0.76},
\end{equation}

where $T_4$ is the gas temperature in units $10^4$ K 
\citep[see, e.g.,][and discussion therein]{richter2008}. 
The temperature term in this equation reflects the temperature 
dependence of the hydrogen recombination coefficient. Since our model provides lower
limits for log $n_{\rm H}$ at a constant value of $T_4=3$, we likewise obtain lower limits
for $f_{\rm HI}$ (Table 4, eigth column) and upper limits for the total hydrogen column
density, $N$(H$)=N($H\,{\sc i}$)+N($H\,{\sc ii}$)=N$(H\,{\sc i}$)/f_{\rm HI}$ (Table 4, 
seventh column). Finally, from the relation $L=N$(H$)/n_{\rm H}$, we get an upper limit 
for the absorption path length $L$ in the absorbers that we can regard as a characteristic
length scale for the absorbing structures. The limits for log $(L/{\rm pc})$ are listed in the 
9th column of Table 4. These span a large range between $1.09$ and $4.83$, corresponding 
to absorber sizes between $12$ pc and $68$ kpc with a median value of $2.8$ kpc.
Similarly large ranges in absorber sizes are 
also found in CGM absorbers, where the ionization conditions can be much more precisely 
constrained using the various transitions from intermediate and high metal ions and 
advanced modeling techniques \citep[e.g.,][]{sameer2024,richter2009}.

\subsection{On the origin and fate of OGrM absorbers}

The absorber sizes, $L$, estimated from our hydrostatic model now can be directly compared 
with the linear distances between neighboring OGrM sightlines, $d_N$ (Sect.\,4.1). 
In the last column of Table 4, we list the ratio (upper limit) between the absorber size and the 
distance to the nearest OGrM absorber, $L/d_N$. Only for four OGrM absorbers in the 
NGC\,5866 group is this ratio clearly $<1$, suggesting that the estimated absorber-size limits are 
(substantially) smaller than the sightline separations. If (regardless of the simplicity of our 
modeling) this estimate reflects reality, these numbers would imply 
that these OGrM Ly\,$\alpha$ absorbers represent individual gas clumps that follow as a 
conglomerate the large-scale motions of matter in the outer regions of the groups' potential wells 
(``swarm-like'', scenario II; see above).

For the remaining majority of the OGrM absorbers, however, the upper limits for $L/d_N$ are all
above unity, some of them substantially. For these systems, both scenarios discussed in 
Sec.\,4.1 are possible and it therefore cannot be excluded that the velocity alignement 
of OGrM absorption along these adjacent sightlines is a result of the fact that
the various LOS that sample a coherent OGrM structure pass through the same kpc-scale 
Ly\,$\alpha$ cloud.

While it is tempting to calculate an upper limit for the absorber's total gas masses
from our simple hydrostatic model to evaluate their contribution to the groups' baryon
budget, we refrain from doing so because of the unknown absorber geometry and the
very large uncertainties that would come along with such an estimate.
Future studies of OGrM absorbers in hydrodynamical cosmological simulations and the analysis of
ultra-deep X-ray observations would be of great help to understand i) whether the high incidence
rate of OGrM absorbers derived in Sect.\,4.2 is a result of AGN feedback, which potentially 
lifts and retains large amounts of warm/hot gas in the outskirts of galaxy groups \citep{eckert2021},
or any other large-scale gas-circulation processes,
ii) whether the $(1-3)R_{\rm vir}$ range around groups hosts a substantial baryon resevoir,
and iii) to what amount the photoionized OGrM absorbers are surrounded by a hotter, collisionally 
ionized medium. Similarly, hydrodynamical simulations of OGrM absorbers in group environments 
of different halo masses and evolutionary states will be important to understand the gas dynamics, 
the importance of OGrM absorbers
for the general properties of the IGrM in the group centers, and their fate in
their role as baryon carrier to feed the disks of group member galaxies and/or their
hot coronal halos with metal-poor gas from the IGM \citep[see][]{afruni2023}.

%%%%%%%%%%%%%%%%%%%%%% SECTION 05 %%%%%%%%%%%%%%%%%%%%%%

\section{Summary and conclusions}

In this paper, we have studied coherent H\,{\sc i} Ly\,$\alpha$ absorption patterns 
in the outer-group medium (OGrM) in the outskirts of four nearby ($D\leq 20$ Mpc), low-mass 
galaxy groups (NGC\,1052 group, NGC\,5866 group, NGC\,4631 group, and NGC\,3992 group)
using archival UV spectral data from HST/COS.
For this, we have analyzed Ly\,$\alpha$ absorption near the groups' systemic velocities 
along 35 AGN sightlines that i) pass the four groups at normalized impact parameters in the 
range $\rho /R_{\rm vir}=1-3$ and ii) are not contaminated by CGM absorption from 
foreground/group-member galaxies. The main results of our study can be summarized as follows:\\
\\
(1) Coherent OGrM H\,{\sc i} Ly\,$\alpha$ absorption (Ly\,$\alpha$ absorption that
is measured at similar group radial velocities along $\geq 2$ adjacent sightlines) 
with equivalent widths between $W_{\lambda}=17$ and $505$ m\AA\, is detected along 19
out of the 35 sightlines. This high detection rate of more than 50 percent indicates 
that diffuse H\,{\sc i}-absorbing gas in the outskirts of galaxy groups has a relatively 
large absorption-cross section.\\ 
\\
(2) From the modeling of the observed Ly\,$\alpha$ absorption profiles we determine
H\,{\sc i} column densities and Doppler parameters for the $19$ OGrM absorbers.
The derived logarithmic H\,{\sc i} column densities lie in the 
range log $N$(H\,{\sc i}$)=12.50-14.34$ with a median value of $13.68$, thus very
similar to the column-density range found in the overall population of Ly\,$\alpha$ forest 
absorbers in nearby ($z=0$) cosmological filaments \citep{bouma2022,wakker2015}, which the OGrM
absorbers are part of. The derived Doppler parameters span a large range ($b=16-80$ km\,s$^{-1}$),
but unresolved velocity structure is likely to be present in many of these lines.
In the impact parameter range sampled by our study ($\rho /R_{\rm vir}=1-3$), 
we do not find any dependency between the Ly\,$\alpha$ equivalent width/H\,{\sc i} column density 
and the group impact parameter. Only one OGrM Ly\,$\alpha$ absorber in our sample 
possibly has associated metal-line absorption (Si\,{\sc iii}), while the vast majority of
the systems are Ly\,$\alpha$-only absorbers.\\
\\
(3) We use a simple hydrostatic toy model for photoionized gas in the outskirts 
of galaxy groups to constrain the physical conditions in the observed OGrM absorber
population. From this, we obtain lower limits for the gas densities from
log $n_{\rm H}=-5.00$ to $-3.72$, comparable to the density range expected
for the Ly\,$\alpha$ forest.
The resulting upper limits for the absorber sizes lie between 12 pc and 68 kpc with a 
median value of $2.8$ kpc. Only for four OGrM absorbers, the estimated size limits 
are smaller than the linear separation between the adjacent OGrM absorbers
in the group, suggesting that these absorbers represent independent gas structures
(tracing gas-density peaks) that move as an ensemble
within or towards the group's potential well. For the remaining 15 absorbers,
no clear statement can be made in this regard.\\
\\
\\
(4) By converting the observed OGrM Ly\,$\alpha$ absorption-cross section into an 
incidence rate, we find that the number of Ly\,$\alpha$ absorbers per unit redshift in nearby 
group environments is $d{\cal N}/dz=232\pm 58$ 
for absorbers with $N$(H\,{\sc i}$)\geq 13.2$ and $\rho /R_{\rm vir}=1-3$.
This is $\sim 25$ percent above the value derived for Ly\,$\alpha$ absorbers
in filaments at $z=0$ \citep[$d{\cal N}/dz=189\pm 25$;][]{bouma2022}
and more than twice the value for the overall $z=0$ Ly\,$\alpha$ forest
\citep[$d{\cal N}/dz=80-100$;][]{danforth2016,lehner2007}
considering the same column-density range. The large incidence rate 
may hint at a substantial reservoir of baryons in the outskirts of galaxy
groups in the form of diffuse ionized gas, but future observations and 
simulations are required to reliably constrain the mass budget in the OGrM and
its connection to AGN feedback \citep{eckert2021}.\\
\\
Our study of OGrM Ly\,$\alpha$ absorbers unveils a number of interesting properties of 
diffuse gas in the outskirts of galaxy groups that deserve to be addressed in a larger
absorption-line survey. We are currently undertaking a fully automated analysis of all 
AGN spectra in the HST/COS archive as part of the QUALSTAR (Quasar Absorption Lines 
Standardized Automated Recovery) project (Richter, in prep.), from which
we will obtain detailed information on $N$, $b$, $z_{\rm abs}$ for several thousand 
Ly\,$\alpha$ absorbers in the nearby Universe. This information will be directly compared
to galaxy data from recent surveys to systematically investigate the connection between 
Ly\,$\alpha$ absorbers and their ambient large-scale environment (e.g., filaments, groups,
and clusters). In this way, we will pinpoint the gaseous matter distribution in the transition zones 
between cosmological filaments and knots in the Cosmic Web. The observational data will
also be compared with mock spectral data synthesized from hydrodynamical cosmological
simulations.

A particularly demanding observational task, yet crucial for a better understanding of gas in group 
environments, will be a reliable estimate of the total gas mass that is hosted in the outskirts 
of groups. Such an estimate can only be obtained from a larger OGrM Ly\,$\alpha$ absorber 
sample in combination with deep X-ray observations that would trace the hot ($T>10^6$ K) phase
of the OGrM which may carry the bulk of the baryons in this region. 
Also, a systematic analysis of the kinematics of Ly\,$\alpha$ absorbers at even
larger distances ($\rho /R_{\rm vir}=1-10$) from the group centers is highly desired to characterize
the streaming motion of these gas absorbers towards the group centers and to estimate the 
contribution of this matter reservoir to the galaxies' gas accretion and star-formation rate.

\begin{acknowledgements}

This research has made use of the SIMBAD database, operated at CDS, Strasbourg, France.
This research also has made use of the NASA/IPAC Extragalactic Database (NED), which is 
funded by the National Aeronautics and Space Administration and operated by the California 
Institute of Technology. The authors would like to thank the anonymous referee for
helpful comments and suggestions.

\end{acknowledgements}

%%%%%%%%%%%%%%%%%%%%%% REFERENCES %%%%%%%%%%%%%%%%%%%%%%

%%%%%%%%%%%%%%%%%%%%%%% APPENDIX %%%%%%%%%%%%%%%%%%%%%%%

\begin{appendix}

\onecolumn

\section{Continuum and absorber modeling}

In Fig.\,A.1, we show as an example the continuum reconstruction and absorption-line
modeling for the OGrM absorber at $v=1457$ km\,s$^{-1}$ in the NGC\,1052 group 
towards the background source NGC\,985 (Table 3). The overall continuum in 
this region between $1218$ and $1226$ \AA\, (red solid line) has been modeled 
from a multi-component Voigt-profile fit of the interstellar H\,{\sc i} Ly\,$\alpha$ 
absorption using the {\tt span} code \citep{richter2013}.
The slow rise of the continuum level reflects the extended Lorentzian wing of the 
interstellar Ly\,$\alpha$ profile. The weak absorber near $1221.5$ \AA\,
is Ly\,$\alpha$ at $v=1457$ km\,s$^{-1}$ in the NGC\,1052 group, which resides
in the wing of the Galactic interstellar Ly\,$\alpha$ absorption. It has been
modeled with a single absorption component with log $N$(H\,{\sc i}$)=12.50$ 
and $b=19$ km\,s$^{-1}$.
All the absorbers listed in Table 3 have been modeled using a similar approach
(see also Sect.\,2.3).

%%%%%%%%%%%%%%%%%%%%%%% FIGURE A1 %%%%%%%%%%%%%%%%%%%%%%%

\begin{figure*}[h!]
\begin{center}
\resizebox{0.7\hsize}{!}{\includegraphics{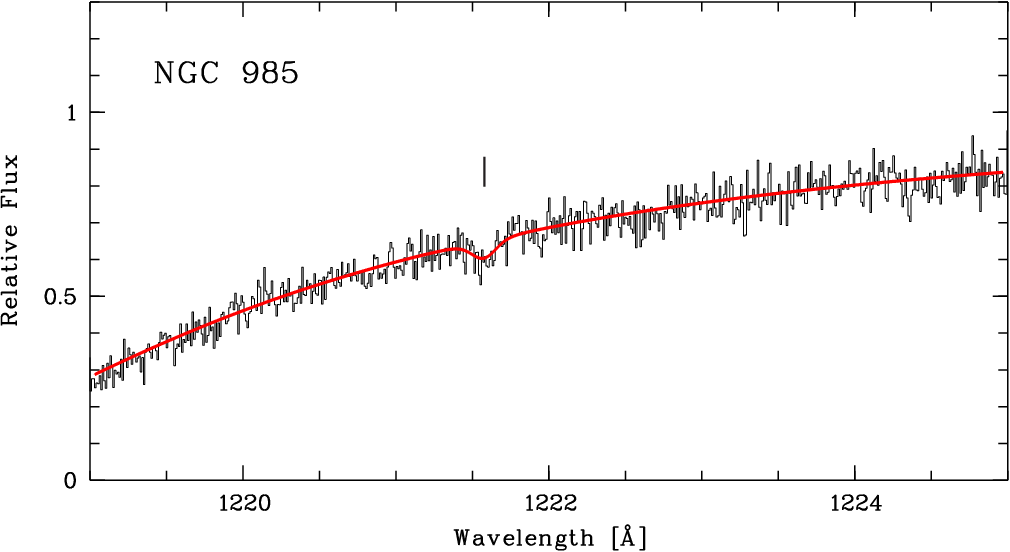}}
\caption[]{
Example for the absorption-line modeling of the OGrM absorber towards NGC\,985 (see also Table 3).
}
\end{center}
\end{figure*}

%%%%%%%%%%%%%%%%%%%%%%%%%%%%%%%%%%%%%%%%%%%%%%%%%%%%%%%%%

\section{Radial density distribution in the OGrM absorbers}

In Fig.\,B.1, we show the logarithmic gas density in the OGrM absorbers 
as a function of the normalized impact parameter to the group center from our 
hydrostatic toy model (Sect.\,4.3). The declining trend for log $n_{\rm H}$ is
a direct result of the assumed hydrostatic equilibrium, with each
group defining its own gravitational potential according to its
mass given in Table 1.

%%%%%%%%%%%%%%%%%%%%%%% FIGURE B1 %%%%%%%%%%%%%%%%%%%%%%%

\begin{figure}[h!]
\begin{center}
\resizebox{0.45\hsize}{!}{\includegraphics{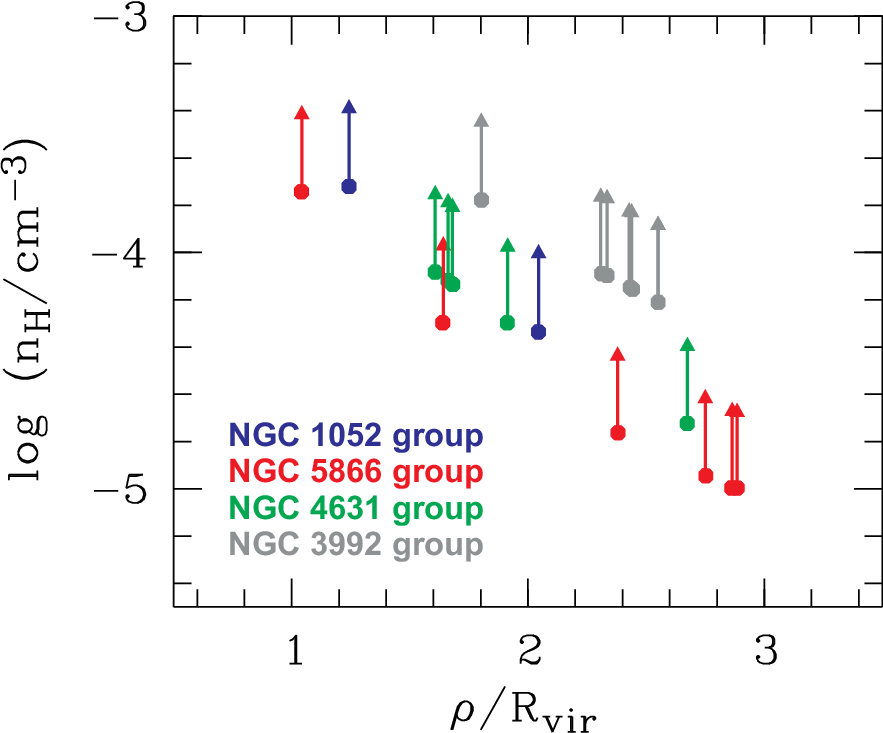}}
\caption[]{
Logarithmic gas density in the OGrM absorbers as a function of
normalized impact parameter from the hydrostatic toy model (Sect.\,4.3).
}
\end{center}
\end{figure}

%%%%%%%%%%%%%%%%%%%%%%%%%%%%%%%%%%%%%%%%%%%%%%%%%%%%%%%%%

\end{appendix}

\end{document}